\documentclass[a4paper,fleqn,usenatbib]{mnras}

\usepackage[T1]{fontenc}
\usepackage{ae,aecompl}

\usepackage{graphicx}	
\usepackage{amsmath}	
\usepackage{amssymb}	
\usepackage{combelow}   

\title[Long-term FRII jet evolution]{Long-term FRII jet evolution in dense environments}

\author[Perucho, Mart\'{\i}, Quilis]{Manel Perucho,$^{1,2}$\thanks{E-mail: manel.perucho@uv.es}
Jos\'e-Mar\'{\i}a Mart\'{\i}$^{1,2}$, Vicent Quilis$^{1,2}$
\\
$^{1}$Departament d'Astronomia i Astrof\'{\i}sica, Universitat de Val\`encia, C/ Dr. Moliner, 50, 46100, Burjassot, Val\`encia, Spain.\\
$^{2}$Observatori Astron\`omic, Universitat de Val\`encia, C/ Catedr\`atic Jos\'e Beltr\'an 2, 46980, Paterna, Val\`encia, Spain.
}

\date{Accepted XXX. Received YYY; in original form ZZZ}

\pubyear{2021}

\begin{document}
\label{firstpage}
\pagerange{\pageref{firstpage}--\pageref{lastpage}}
\maketitle

\begin{abstract}
We present long-term numerical three-dimensional simulations of a relativistic outﬂow propagating through a galactic ambient medium and environment, up to distances $\sim 100$~kpc. Our aim is to study the role of dense media in the global dynamics of the radio source. We use a relativistic gas equation of state, and a basic description of thermal cooling terms. In previous work, we showed that a linear perturbation could enhance the jet propagation during the early phases of evolution, by introducing obliquity to the jet reverse shock. Here, we show that this effect is reduced in denser media. We find that the \emph{dentist-drill} effect acts earlier, due to slower jet propagation and an increased growth of the helical instability. The global morphology of the jet is less elongated, with more prominent lobes. 
The fundamental physical parameters of the jet generated structure derived from our simulations fall within the estimated values derived for FRII jets in the 3C sample. In agreement with previous axisymmetric and three dimensional simulations in lower density media, we conclude that shock heating of the interstellar and intergalactic media is very eﬃcient in the case of powerful, relativistic jets.
\end{abstract}

\begin{keywords}
Galaxies: active  ---  Galaxies: jets --- Hydrodynamics --- Shock-waves --- Relativistic processes --- X-rays: galaxies: clusters
\end{keywords}



\section{Introduction}

Powerful extragalactic jets are generated in the environment of supermassive black holes, in Active Galactic Nuclei (AGN), as a result of accretion of matter onto these objects. The Blandford-Znajek mechanism of extraction of rotational energy from the black hole \citep{bz77} represents the most plausible option of jet generation, as supported by numerical simulations \citep[e.g.,][]{tch11,bk12,po13,tch15}. Recent works have allowed radio astronomers to map regions closer to these objects \citep{ha11,ha16,ba16,eht19a,bo21}, showing that jets collimate within the inner $10^3-10^6$ gravitational radii. Once ejected and accelerated to relativistic speeds by thermal and magnetic mechanisms \citep{li09,ho15}, jets propagate through hundreds of kiloparsecs. Acceleration of these outflows to super(magneto)sonic speeds causes shocks that cross the host galaxy and beyond \citep[e.g.,][for observational evidence and numerical simulations]{nu05,mc05,si09b,git10,pqm11,cro11,sta14,pmqr14,pmq19,mm19,ro20,mu20,seo21}. 

At kiloparsec scales, extragalactic jets are mainly observed at radio frequencies, and show a morphological and brightness dichotomy between powerful sources showing bright hotspots and radio lobes, and less powerful radio sources that show irregular, decollimated structures, which was first noted by \cite{fr74}. The main reason responsible for this dichotomy is probably jet power \citep{rs91,gc01}, with a transition region in which the environmental properties and stellar populations may play a significant role \citep[e.g.,][]{ka09,pe19,pe20,ac21}. 

Numerical simulations have shown to be useful tools to reproduce the gross large-scale features of powerful relativistic flows, namely, the presence of bow-shocks, hot-spots and lobes in radio galaxies. Nevertheless, key parameters such as the age of the sources, which are used in the literature to evaluate the jet power, for instance, are only estimated making approximations to spectral ages \citep[e.g.,][]{mu99,ma07,od09,har16}.   

In a previous paper \citep[][Paper I from now on]{pmq19}, we have studied the long-term evolution of a powerful, FRII jet in three dimensions, evolving through the hot component of the interstellar and intergalactic media (ISM and IGM, respectively). This media are described through density and pressure profiles in hydrostatic equilibrium, based on X-ray observations of the radio-galaxy 3C31 \citep{hr02}. The simulation showed fast jet propagation as compared to previous two-dimensional models with the same set up, which was assigned to the linear growth of a helical instability at the jet head, resulting in obliquity at the terminal shock and favouring the acceleration across the galactic atmosphere. Once the oscillation becomes non-linear, the {\it dentist-drill effect} \citep{sch74} contributes to decelerating the jet advance. The numerical resolution of the jet was insufficient to properly study jet dynamics, because the work was originally designed to probe the large-scale structure. However, we showed that an increase of the resolution enhanced this effect, as expected from faster mode amplitude growth with increased resolution (and the consequent drop of numerical viscosity). Altogether, we found that this head-wobbling can generate a well collimated structure with elongated lobes resembling the morphology of a number of FRII jets (e.g., 3C~33, 3C~46, 3C~219, 3C~273; see Paper~I). However, many other FRII radio galaxies show thicker lobes and their morphology cannot thus be explained in the same terms. In morphological terms, the simulations run by \citet{ma19} gave very similar results to those reported in Paper~I, although in this paper the authors use a larger number of cells per jet radius -at the cost of a large jet radius ($R_j=2\,{\rm kpc}$, with 0.4~kpc per cell, versus 0.1~kpc per cell in Paper~I) and a smaller grid.

In this paper, we present results of a long-term 3D numerical simulation in which we have used the same setup as that in Paper~I for the jet (jet kinetic power, $L_k=10^{45}\,{\rm erg/s}$), with an increased ambient medium density and pressure by a factor 4. In this way, we pretend to study the large-scale structures generated by jets in denser galactic clusters, as those surrounding radio sources as 3C~228, 3C~244.1, 3C~267, or 3C~427.1, with estimated IGM densities $\sim 10^{-27}\,{\rm g\,cm^{-3}}$ or pressures $\sim 10^{-11}\,{\rm dyn\,cm^{-2}}$, i.e., close to an order of magnitude above the values considered in previous simulations \citep[as derived for the radio galaxy 3C31 by][]{hr02}. We have also included radiative (thermal) cooling terms in our equations, with the aim to study the possible effects of radiation on the shock properties at large scales. We compare the global jet thermodynamics with previous simulations and study the impact of the jet on the IGM. 

The paper is structured as follows. The setup of the simulations is presented in Section~2. Results are given in Section~3. Section~4 includes the discussion and conclusions of this work.

\section{Numerical simulations} \label{sec:setup}

\subsection{The code} \label{ss:code}

We have used the code \emph{Ratpenat}, a hybrid parallel code  -- MPI + OpenMP -- that solves the  equations of relativistic hydrodynamics in conservation form, using high-resolution-shock-capturing methods \citep[see][Paper~I, and references therein]{pe10}. We have also used IDL software and LLNL VisIt \citep{visit} to produce the figures. The accurate description of the equations that we solve is presented in Paper~I. In this work, we have added thermal cooling as a source term, $\Lambda$, in the energy equation \citep[see also][]{pbrb17}. The idea is to check the role of bremsstrahlung on the shock properties at the scales considered. The cooling term is defined (in cgs units) as in the approximation given in \cite{mya98} (where $T$ is the gas temperature):

\begin{equation}
 \Lambda\,=\, n_e\,n_Z \times
 \left\{ 
 \begin{array}{lr}
 7\times10^{-27}T,& 10^4\leq T \leq 10^5  \\  \nonumber
 7\times10^{-19}T^{-0.6},& 10^5\leq T \leq 4\times10^7 \\
 3\times10^{-27}T^{0.5},& T \geq 4\times10^7    \nonumber
 \end{array} \right\} \,
\end{equation}

\noindent
where $n_e = \rho_e/m_e$ and $n_Z=n_p =\rho_p/m_p$ ($Z=1$, we assume ionized hydrogen alone for simplicity) are the electron and ion number densities. The Synge equation of state \citep[][described in Appendix A of \citealt{pm07}]{sy57} that allows us to describe a mixture of relativistic ideal gases, provides those parameters. Thus, we implicitly assume thermodynamical equilibrium at each cell. Finally, we also assume that these losses need only to be accounted for the pre and post-shock ambient medium, where $v\ll c$. This implies that the losses are isotropically radiated (the medium is dilute and thus optically thin) and cooling only needs to be considered in the energy equation \citep[see][]{pbrb17}.      

\subsection{Set up} 
\label{sec:su}

The setup of the 3D simulations presented in this paper is the same as in Paper~I, but for the aforementioned increase in the ambient density and pressure. The 3D grid is filled by the ambient gas with a density and temperature profiles equivalent to those used in previous papers, as derived from the modellisation of the X-ray observations of the radio galaxy 3C~31 \citep{hr02}: 
\begin{eqnarray}\label{next}
  n_{\rm ext} = n_{\rm c} \left(1 +
\left(\frac{r}{r_{\rm c}}\right)^2\right)^{-3\beta_{\rm atm,c}/2} +  \nonumber \\
+ n_{\rm g} \left(1 + \left(\frac{r}{r_{\rm g}}\right)^2\right)^{-3\beta_{\rm atm,g}/2},
\end{eqnarray}
where $r$ is the radial spherical coordinate. Here, we have used $n_{\rm c} = 0.72$~cm$^{-3}$ and $n_{\rm g} =0.0076$~cm$^{-3}$, i.e., we have increased the ambient medium density by a factor of four with respect to Paper~I. We keep $r_{\rm c} = 1.2$~kpc, $\beta_{\rm atm,c} = 0.73$ for the galaxy, and $r_{\rm g} = 52$~kpc and  $\beta_{\rm atm,g} = 0.38$ for the group density distribution. The temperature profile is the same as in Paper~I:

\begin{equation}\label{text}
T_{\rm ext} = \left\{ \begin{array}{ll}
T_{\rm c} + (T_{\rm g} - T_{\rm c}) \displaystyle{\frac{r}{r_{\rm m}}}, & {\rm for\,} r\leq r_{\rm m} \\
T_{\rm g}, &  {\rm for\,} r > r_{\rm m}
\end{array} \right.
\end{equation}
where $T_{\rm c}$ and $T_{\rm g}$ are characteristic temperatures of the host galaxy and the group ($4.9 \times 10^6$ K and $1.7 \times 10^7$ K, respectively), and $r_{\rm m} = 7.8$~kpc is the matching radius between the cool-core and the surrounding, almost isothermal IGM. With these conditions, the external pressure is derived from the number density and temperature profiles assuming a perfect gas composed of ionized hydrogen \citep{hr02,pm07}:
\begin{equation}\label{pext}
  p_{\rm ext} = \frac{k_{\rm B} T_{\rm ext}}{\mu X} n_{\rm ext},
\end{equation}
where $\mu$ is the mass per particle in atomic mass units ($\mu=0.5$ here), $X$ is the abundance of hydrogen per mass unit, which is set to 1, and $k_{\rm B}$ is the Boltzmann's constant. Therefore, the increase in $n_{\rm ext}$ translates into an equivalent increase in $p_{\rm ext}$. The ambient medium is kept in equilibrium by means of a restoring force (an external gravity) as described in Paper~I. All these parameters represent a moderate size galaxy cluster with mass $4\times10^{14}\,M_{\odot}$ and $\sim 1\, \rm{Mpc}$ virial radius. 

Regarding the jet parameters at injection (1~kpc from the active nucleus), they are also the same as in Paper~I, namely, kinetic power $L_{\rm k}=10^{45}\,{\rm erg\,s^{-1}}$, jet radius $R_{\rm j}=100$~pc, flow velocity at injection $v_{\rm j}=0.984\,c$, specific internal energy $2.5\times10^{-3}\,c^2$, and density ratio between the jet material and environment of $\rho_{\rm j}/\rho_{\rm a,0} = 1.25\times 10^{-4}$ (four times less than in Paper~I), resulting in $\rho_{\rm j}=8.3\times 10^{-29}$~g/cm$^3$ ($\rho_{\rm a,0} = 6\times10^{-25}$~g/cm$^3$), with a purely leptonic jet composition. We have also introduced the oscillatory perturbations in the flow at injection, as in Paper~I, with frequencies $w_1=0.01\,c/R_{\rm j}$, $w_2=0.05 \,c/{\rm R_{\rm j}}$, $w_3=0.1\,c/R_{\rm j}$, and $w_4=0.5\,c/R_{\rm j}$, in order to trigger the development of helical structures with different wavelengths:          
 
\begin{eqnarray}
   v_{x} = 2.5\times10^{-4} \, v_{\rm j} \sum_{i=1}^{4} \cos(w_i\,t),
   \nonumber \\
   v_{z} = 2.5\times10^{-4} \, v_{\rm j} \sum_{i=1}^{4} \sin(w_i\,t).
\end{eqnarray}

The numerical grid involves $1024\times2048\times1024$ cells in the $x$, $y$, and $z$ coordinates, respectively, where $y$ is the direction of jet propagation. The resolution is $(100~{\rm pc})^3$ per cell (i.e., 1 cell per jet radius at injection). The limited resolution is justified in terms of our interest on global parameters and jet generated structure, and because the expected jet expansion increases this resolution with distance, but see Paper~I for a discussion on resolution. The grid therefore has a physical size $\simeq 100\times200\times100$~kpc. 

The boundary conditions are outflow at all boundaries, except at the jet base, where we programmed injection conditions at the jet inlet, and reflecting conditions for the rest of this plane. Reflection is typically used in order to avoid a drop in pressure if backflow reaches this limit of the grid, and to simulate the effect of the presence of a counter-jet. The simulations were run in Mare Nostrum, at the Barcelona Supercomputing Centre and in Tirant, at the Servei d'Inform\`atica de la Universitat de Val\`encia, where we also performed the analysis of the results, with a hybrid MPI/OMP program, \emph{Mussol}.   

In this paper, we will discuss the results of simulation J1 as compared to J0 (the simulation in Paper~I). Because the lack of a priori knowledge on the time-scales involved in the simulation, we also ran a second simulation, J1b, without cooling, in order to check possible large-scale effects of cooling in the simulation. The time-scales involved were, nevertheless, too short for this process to play a major role and the results did not introduce any significant difference in the large-scale structure and evolution. Therefore, we do not present this second simulation here.

\section{Results} 
\label{sec:res}
\subsection{Kinematics}

%
\begin{figure*}  
\includegraphics[trim=3cm 13cm 0 4cm,width=0.48\textwidth]{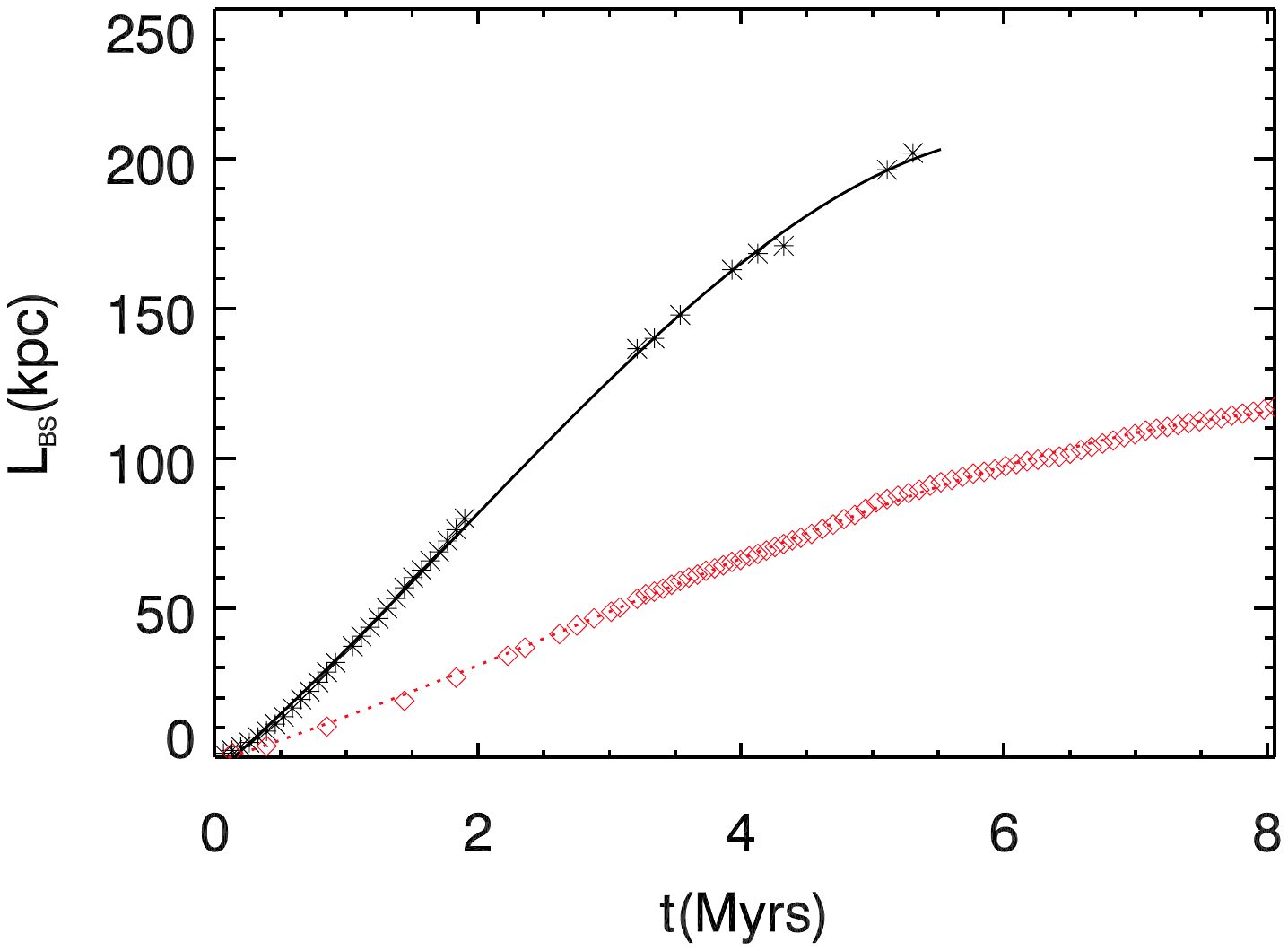}
\includegraphics[trim=3cm 13cm 0 4cm,width=0.48\textwidth]{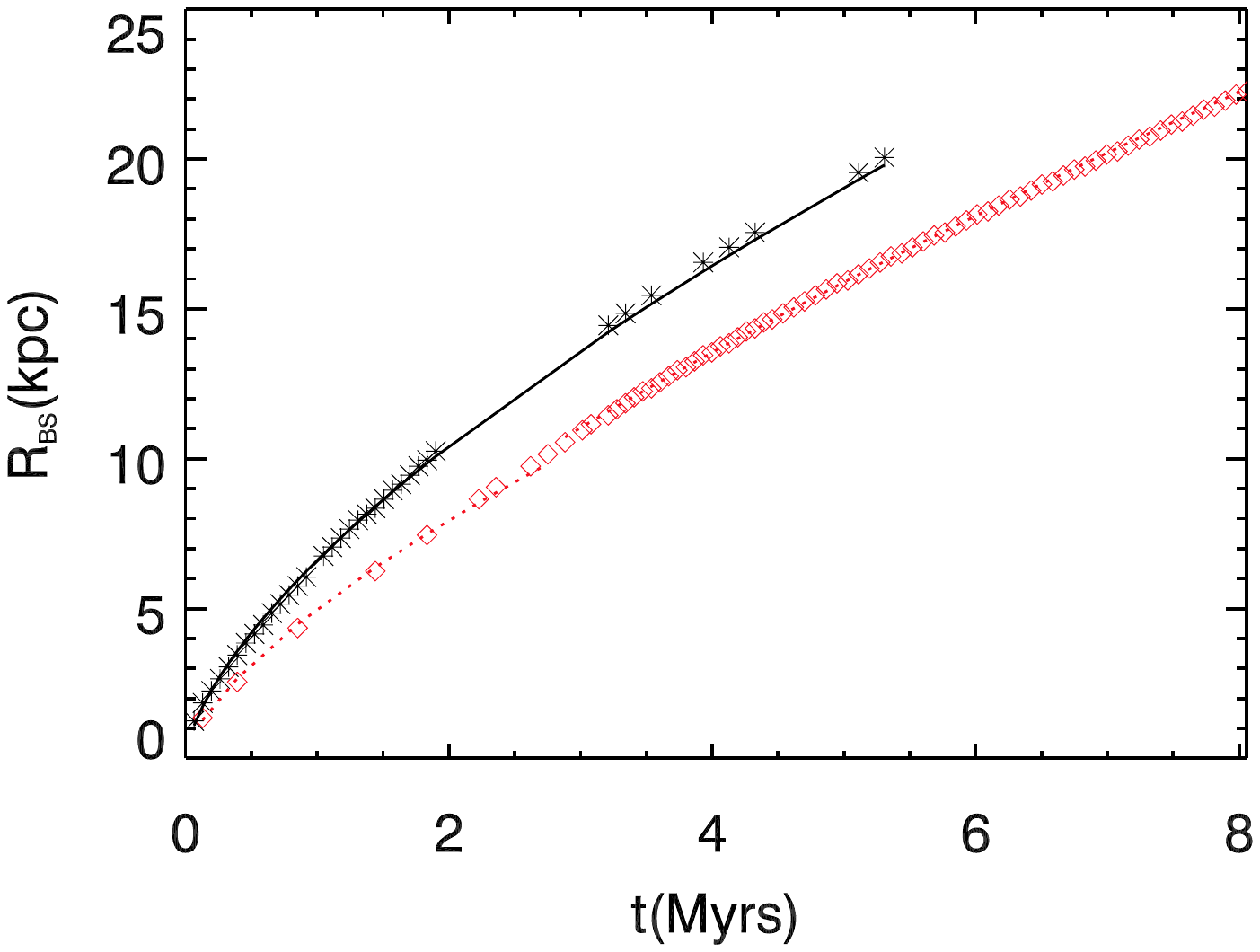}
\caption{Time evolution of the axial bow-shock position (left),
and cocoon maximum radius (right). The asterisks stand for simulation J0 (Paper~I), the red rhombi indicate the simulation J1. The lines show the fits to the evolution of radius and head position with time (see the text).}
\label{fig:kine}
\end{figure*}
%

Figure~\ref{fig:kine} shows the evolution of the shocked region for the different jet simulations. The left panel displays the position of the head of the jets (approximated here by the most advanced position of the bow shock along the propagation axis) as a function of time. The propagation speed of model J1 is smaller than that of model J0 as expected from its denser environment. 

A polynomial fit of third order of the jet head position as a function of time has been performed, showing a positive coefficient for the second degree term and a negative coefficient for the third degree one, $L_{\rm BS}[{\rm kpc}] = a_0 +a_1 t[{\rm Myr}] + a_2 t[{\rm Myr}]^2 + a_3 t[{\rm Myr}]^3$, with $a_0 = 1.45 \pm 0.52$, $a_1 = 13.95 \pm 0.47$, $a_2 = 1.43 \pm 0.13$, and $a_3 = - 0.17 \pm 0.01$ (the errors are given by the {\it poly\_fit} IDL routine). This result shows that the data reproduce an axial propagation of the jet with a varying (first positive, then negative) acceleration. The break between these two behaviours occurs at $\sim 2.8$ Myr. Thus, we take this value as the reference time to separate the evolution between the initial and  multidimensional phases to be discussed in Section~\ref{ss:jd}. Compared with the simulations of Paper~I (here represented by model J0) the initial acceleration phase across the density drop is longer (2.8 Myr instead of 2.0 Myr) but less pronounced ( $\sim 1.5$ kpc/Myr$^2$ instead of $\sim 5$ kpc/Myr$^2$). 
In Paper~I we showed that, besides the density gradient in the ambient medium, a second factor favouring the acceleration of the jet was the development of instabilities that forced small-amplitude oscillations at the jet head, resulting into an oblique terminal shock. Since only the normal component of the flow velocity is decelerated at a shock, this wobbling at the head of the jet allows for a faster propagation than in the case of a normal terminal shock \citep{alo99}. This phase of efficient jet propagation ends when the oscillation at the jet head becomes nonlinear and the {\it dentist-drill effect} \citep{sch74} starts dominating the evolution. As discussed in Paper~I, simulations with increased resolution showed enhanced jet acceleration because of a faster development of instabilities as resolution is increased (and numerical viscosity is reduced), but also a faster transition to the nonlinear phase. Altogether, the increased resolution implies faster propagation during this period, which is shorter, so the jet reaches longer distances in shorter times before transiting into the {\it dentist-drill} phase. A more detailed discussion about the jet propagation speed will be presented in Section~\ref{ss:jd}.

The right panel of Figure~\ref{fig:kine} shows that the radius of the shocked region is also smaller in model J1. As a result, the volume embedded in the shocked jet region is also larger in J0. This result reflects again the larger opposition of the ambient medium in simulation J1 to the jet expansion. 

%
\begin{figure*}
\includegraphics[trim=3cm 13cm 0 4cm,width=0.48\textwidth]{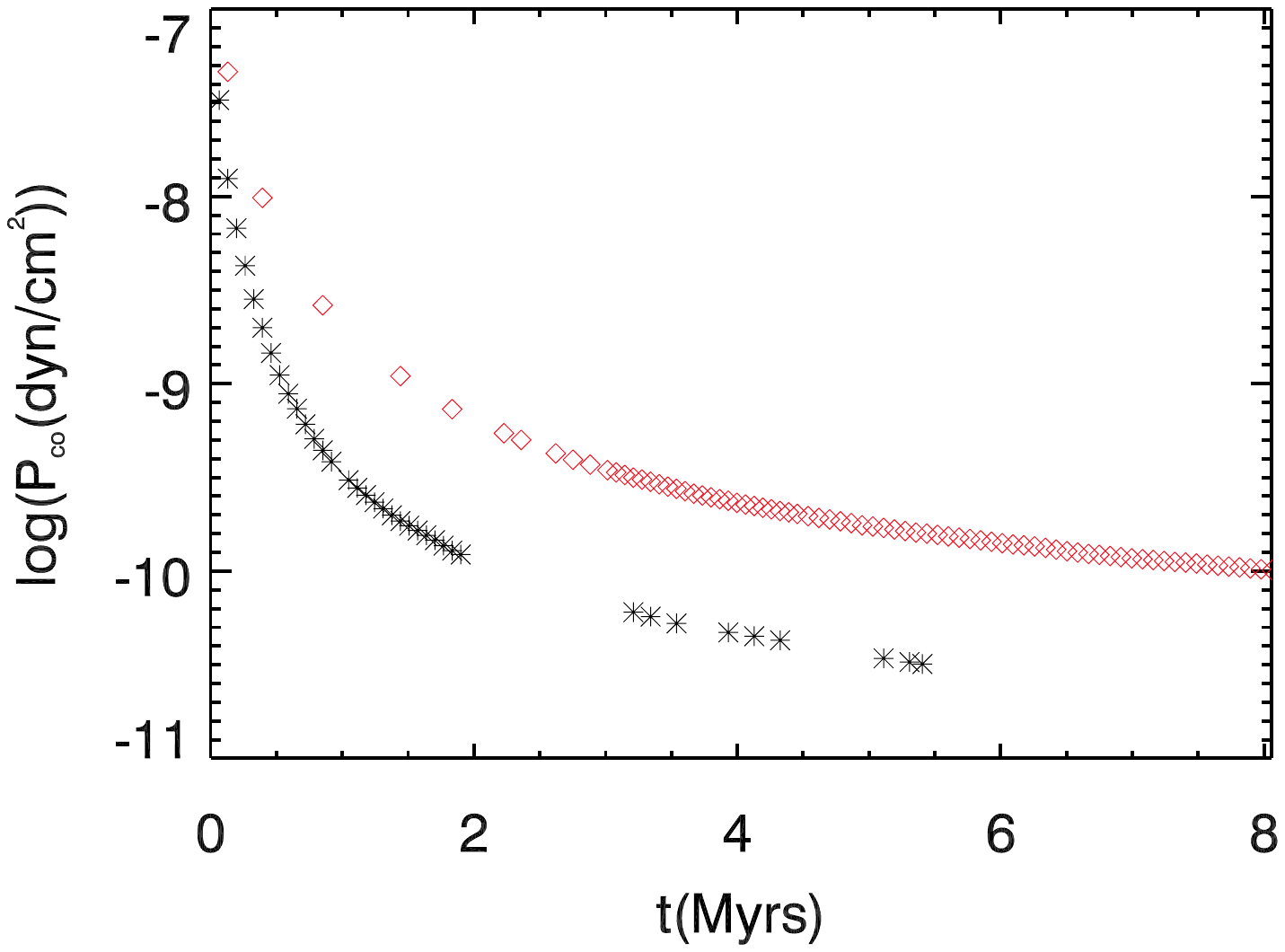}
\includegraphics[trim=3cm 13cm 0 4cm,width=0.48\textwidth]{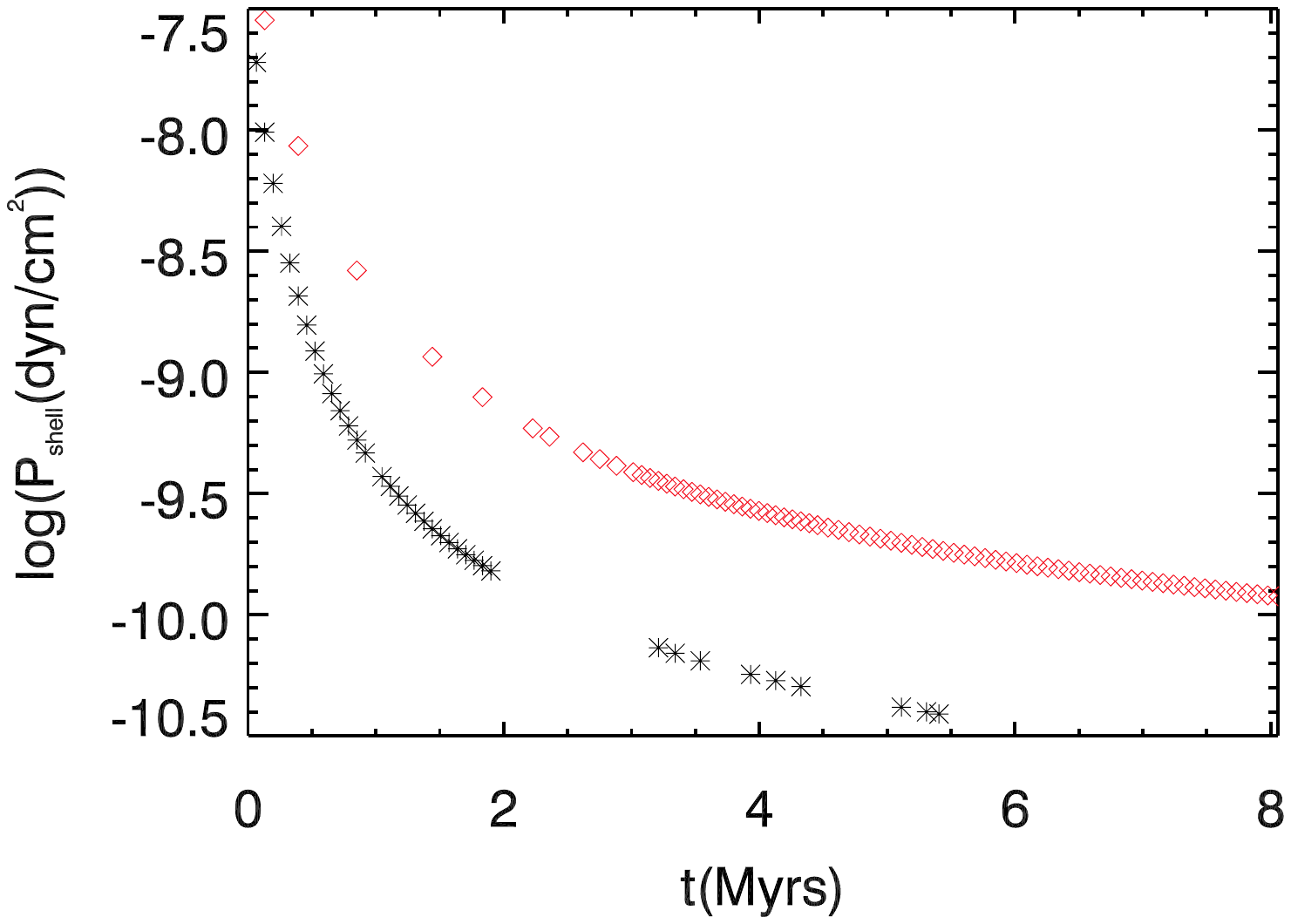}
\includegraphics[trim=3cm 13cm 0 4cm,width=0.48\textwidth]{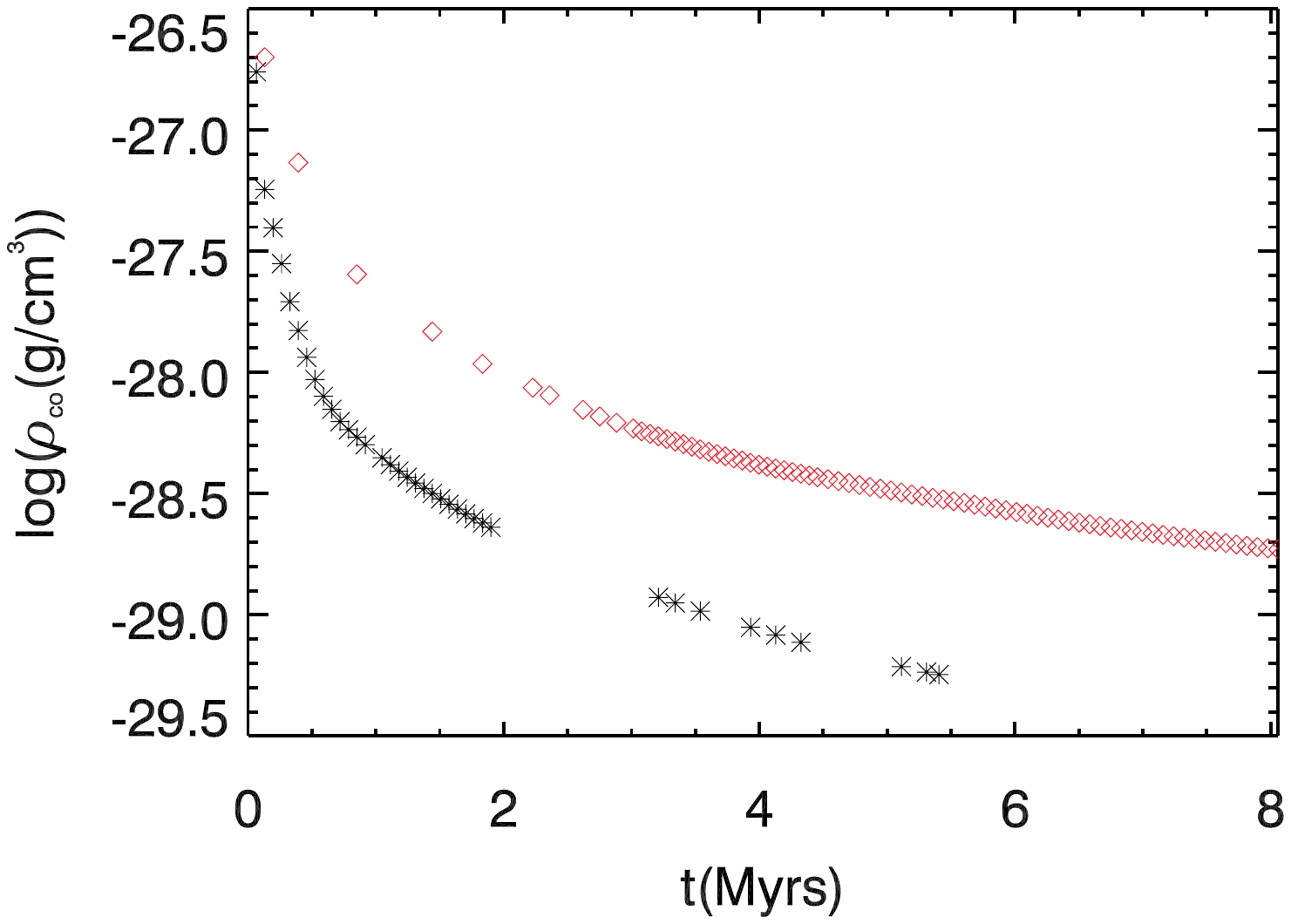}
\includegraphics[trim=3cm 13cm 0 4cm,width=0.48\textwidth]{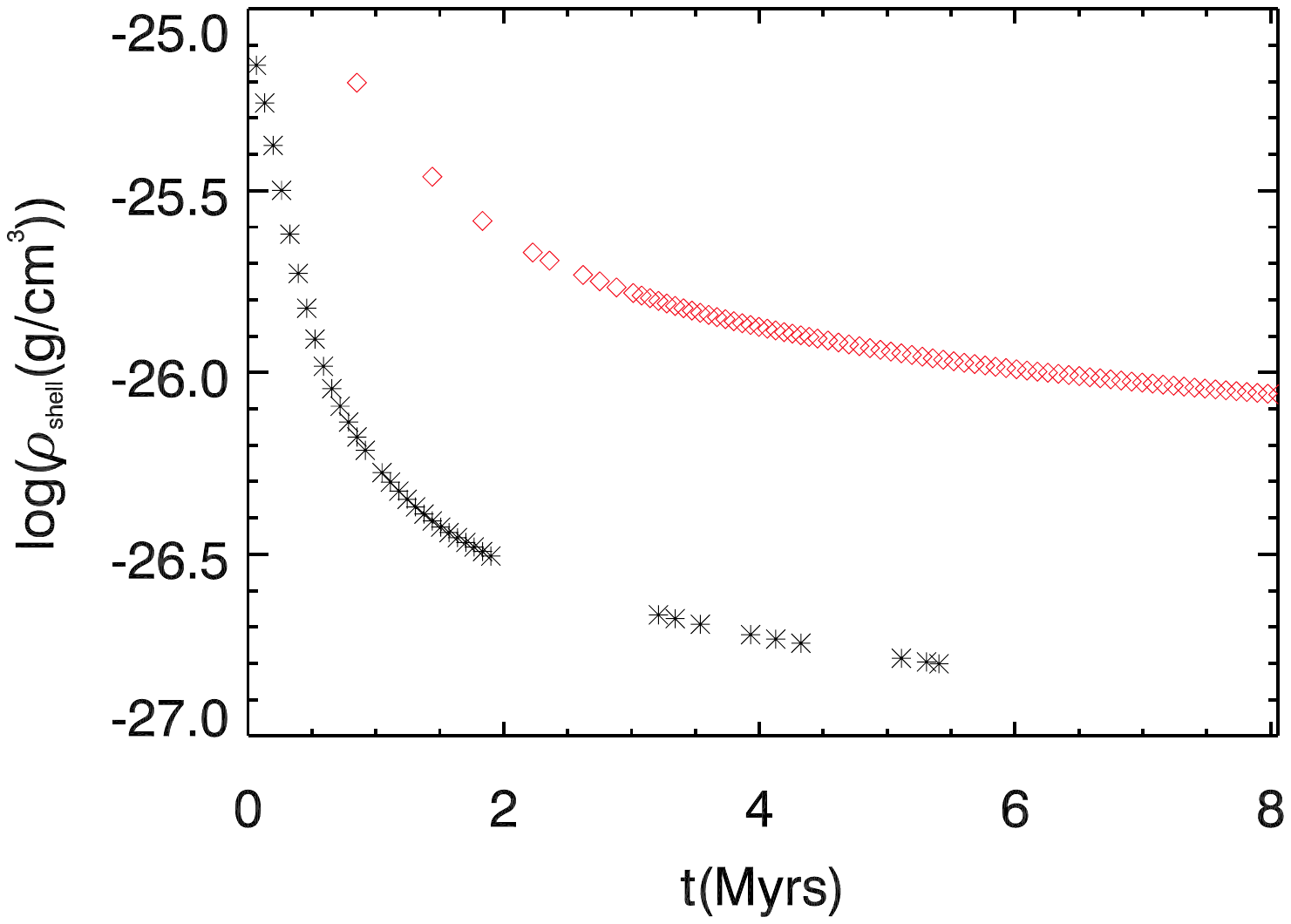}
\includegraphics[trim=3cm 13cm 0 4cm,width=0.48\textwidth]{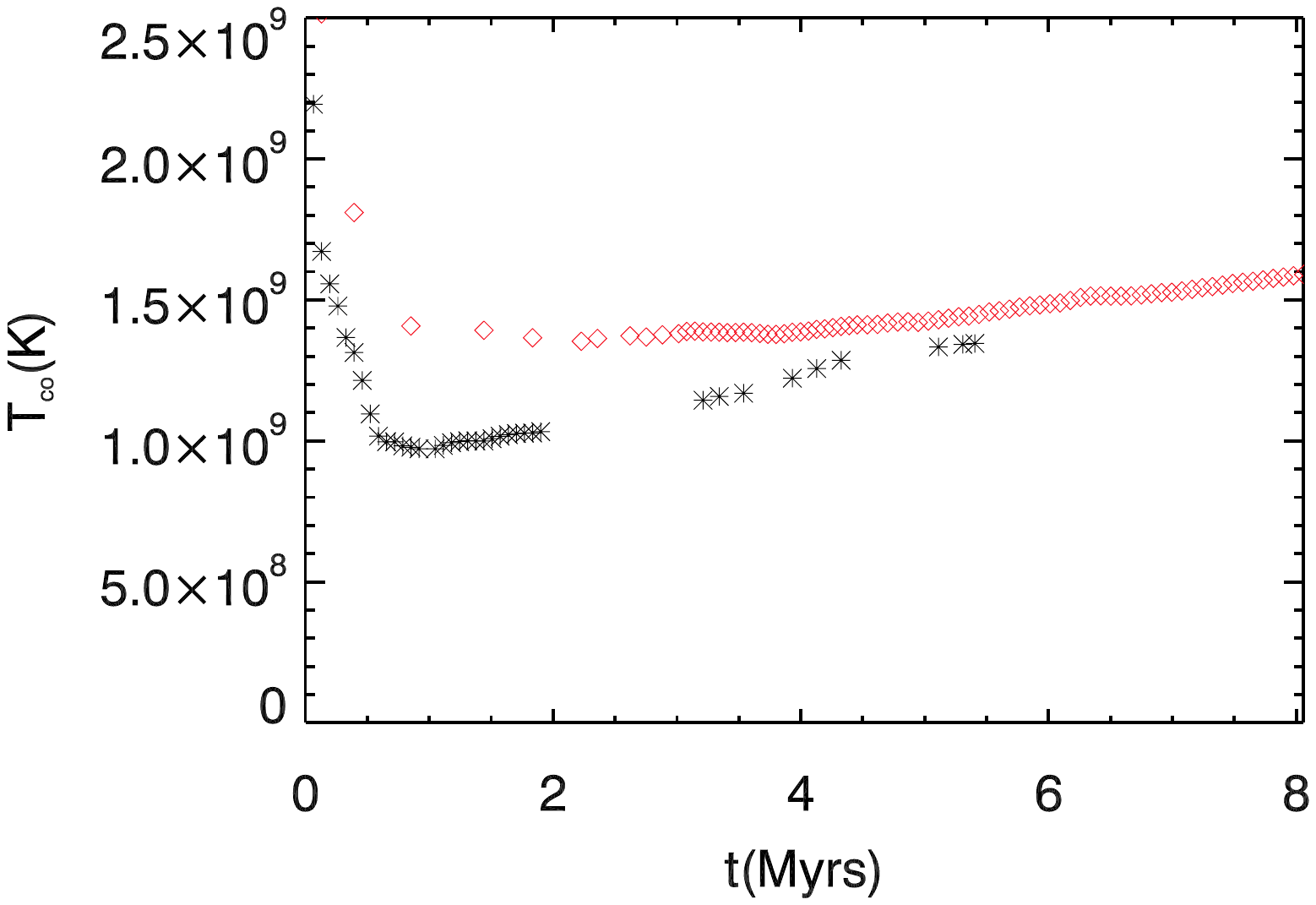}
\includegraphics[trim=3cm 13cm 0 4cm,width=0.48\textwidth]{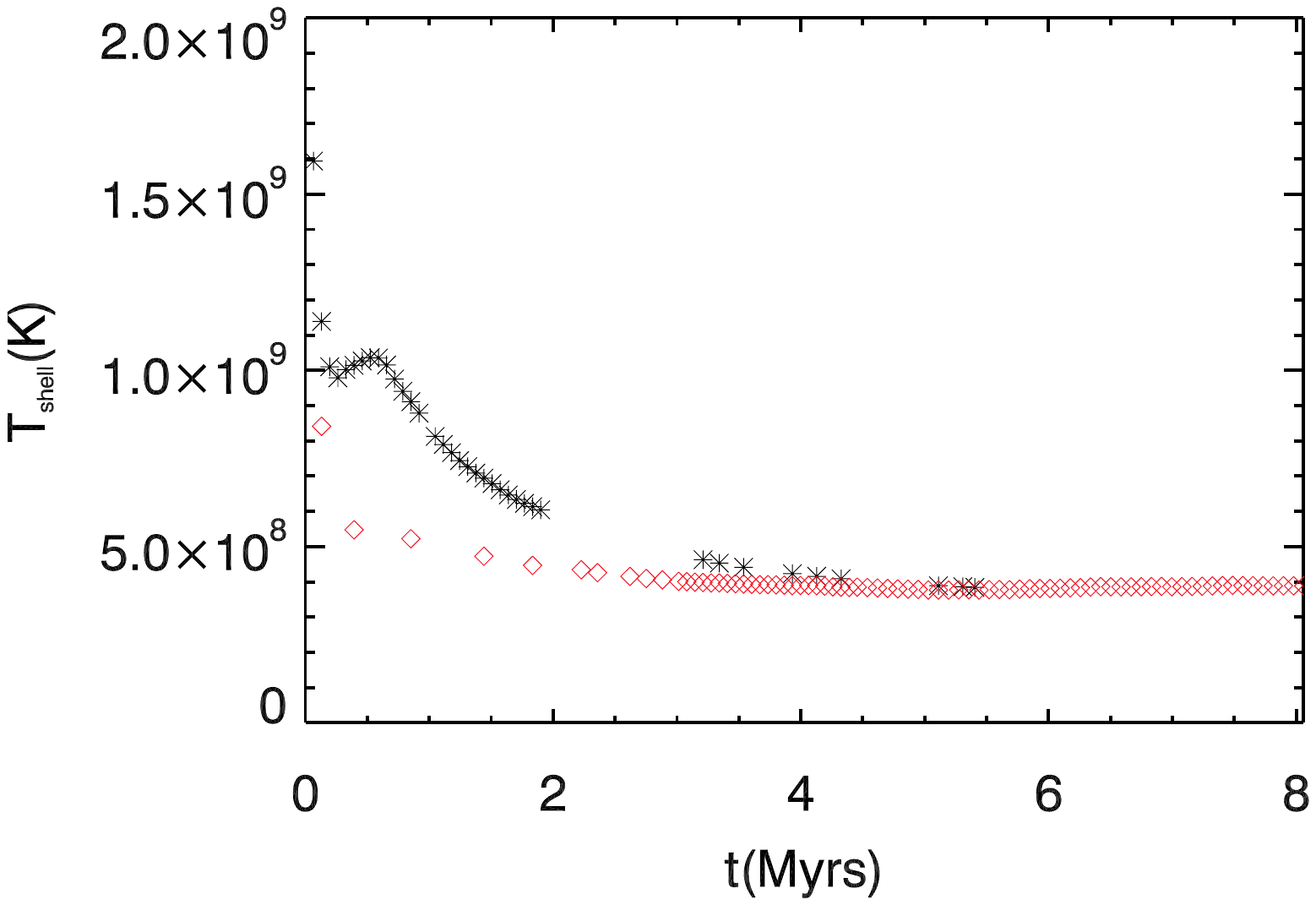}
\caption{Mean values of pressure (top), density (centre) and temperature (bottom) in the cocoon (left column) and shocked ambient medium (shell, right column). The asterisks stand for simulation J0 (Paper~I), and the red rhombi indicate the simulation J1.}
\label{fig:prhot}
\end{figure*}
%

%
\begin{figure} 
\includegraphics[trim=3cm 13cm 0 4cm,width=0.5\textwidth]{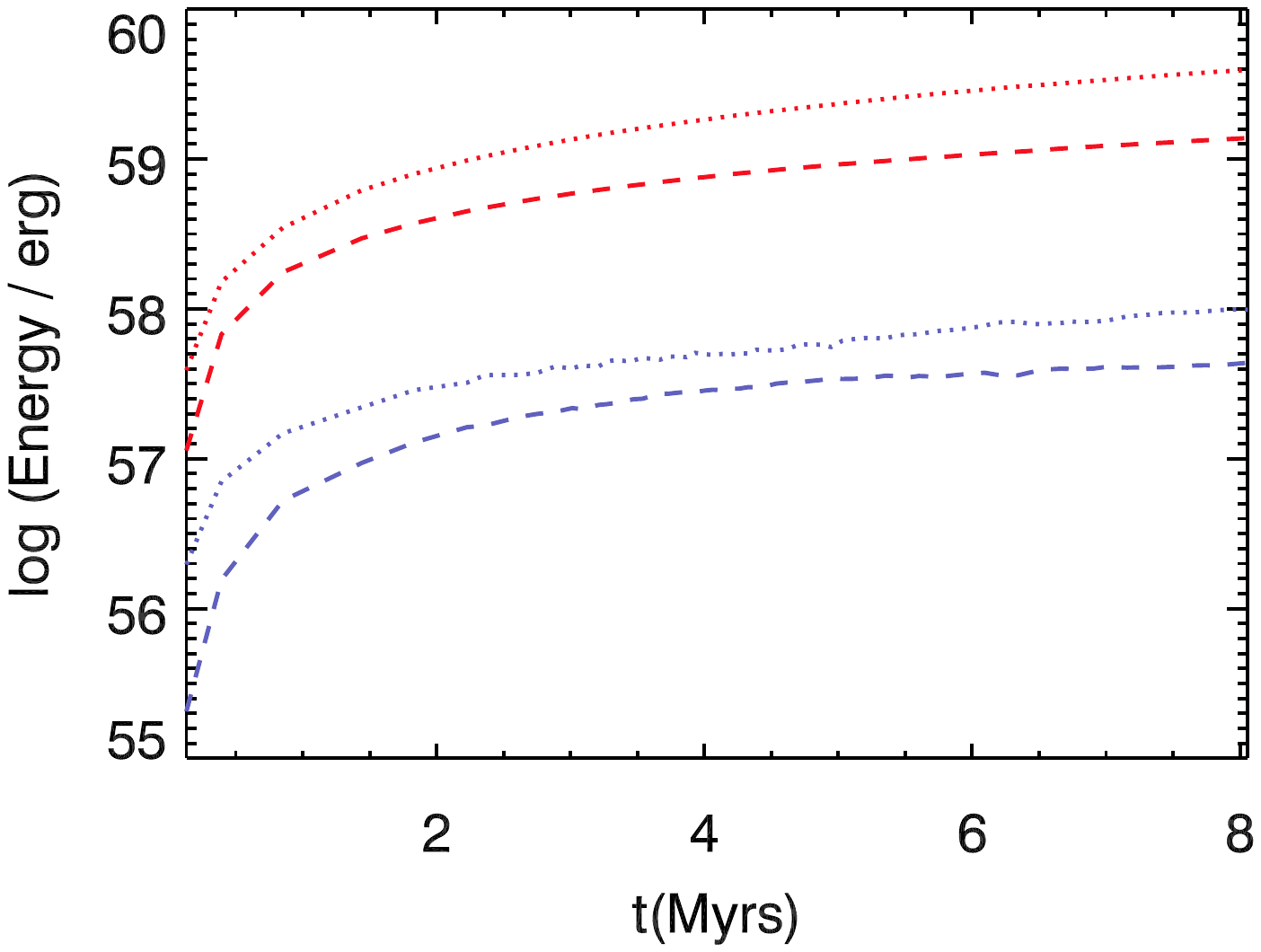}
\caption{Time evolution of the logarithm of the energy for simulation J1. The red dotted (dashed) line represents the increase of internal (kinetic) energy in the processed ambient medium. The blue dotted (dashed) line displays the internal (kinetic) energy of the shocked jet material (i.e., cocoon).}  
\label{fig:ene}
\end{figure}
%

\subsection{Thermodynamics}
Figure~\ref{fig:prhot} displays the evolution of pressure (top panels), rest-mass density (centre panels) and temperature (bottom panels) for the cocoon (shocked jet gas, left) and shell (shocked IGM, right). On the one hand, from the evolution of models J1 and J1b (a simulation with no cooling, not shown here), we observe that cooling does not play any role in the evolution of the global thermodynamical variables, even if it could be locally relevant at some points during the jet evolution. On the other hand, the increased environmental density manifests in larger values of cocoon and shocked IGM densities and pressures in simulations J1a and J1b with respect to J0. Finally, both density and pressure adjust in such a way that, as justified in \citet{pm07}, the temperature of these regions remains fairly constant along the evolution (as long as the injection fluxes through the jet terminal shock remain constant).

Figure~\ref{fig:ene} shows the time evolution of the 
internal and kinetic energy gained by the ambient medium (red dotted and dashed lines, respectively), and the internal and kinetic energy kept by the jet plasma (blue dotted and dashed lines, respectively). These energies are computed with the expressions defined in Paper~I. The evolution and the fractions at the end of the simulation of the different kinds of energy (74\% transferred to the ambient medium as internal energy;  24\% transferred to the ambient medium as kinetic energy; 2\% stored in the jet plasma) are very similar to those obtained in previous works \citep{pqm11,pmqr14,pmq19}. Again, this shows that the shock heating is a highly efficient mechanism for the transfer of energy to the ambient medium. This result represents one further evidence that collimated outflows transfer a large fraction of the injected power to the ambient medium via strong shocks and that it should be taken into account in cosmological models including AGN feedback \citep[see also][]{pmqb17}. 

%
\begin{figure} 
\includegraphics[width=0.45\textwidth]{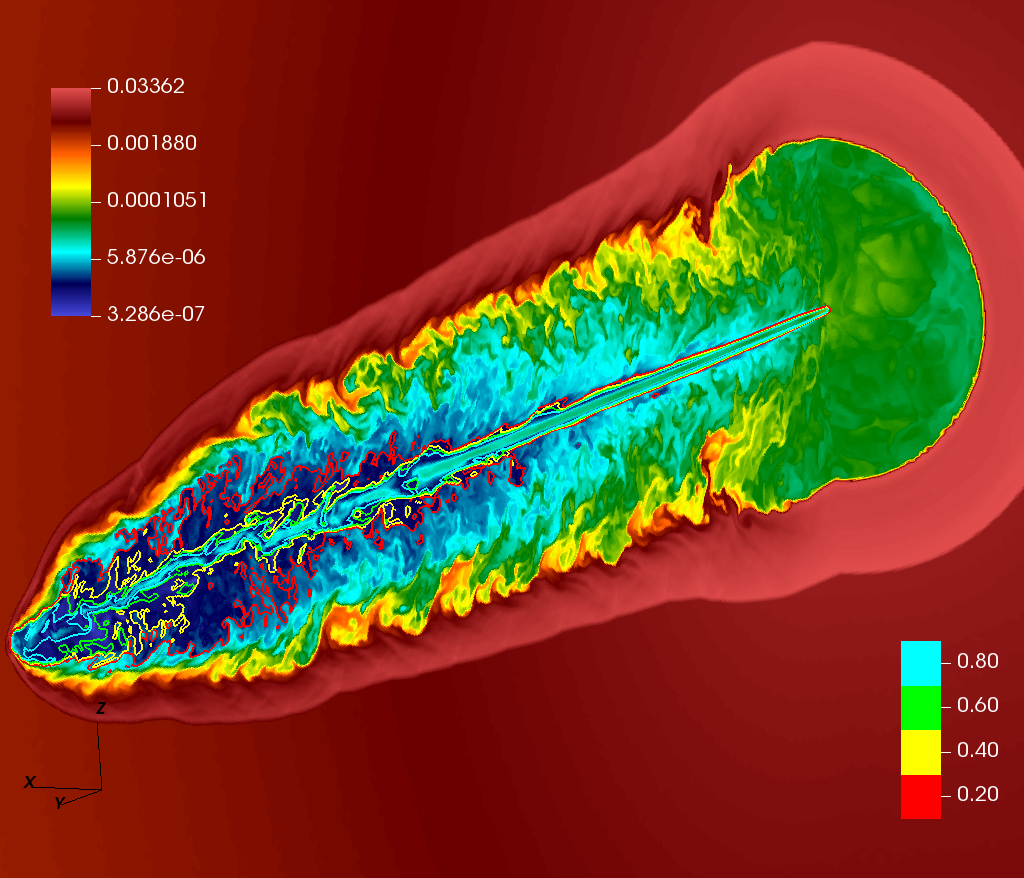}\\ \includegraphics[width=0.45\textwidth]{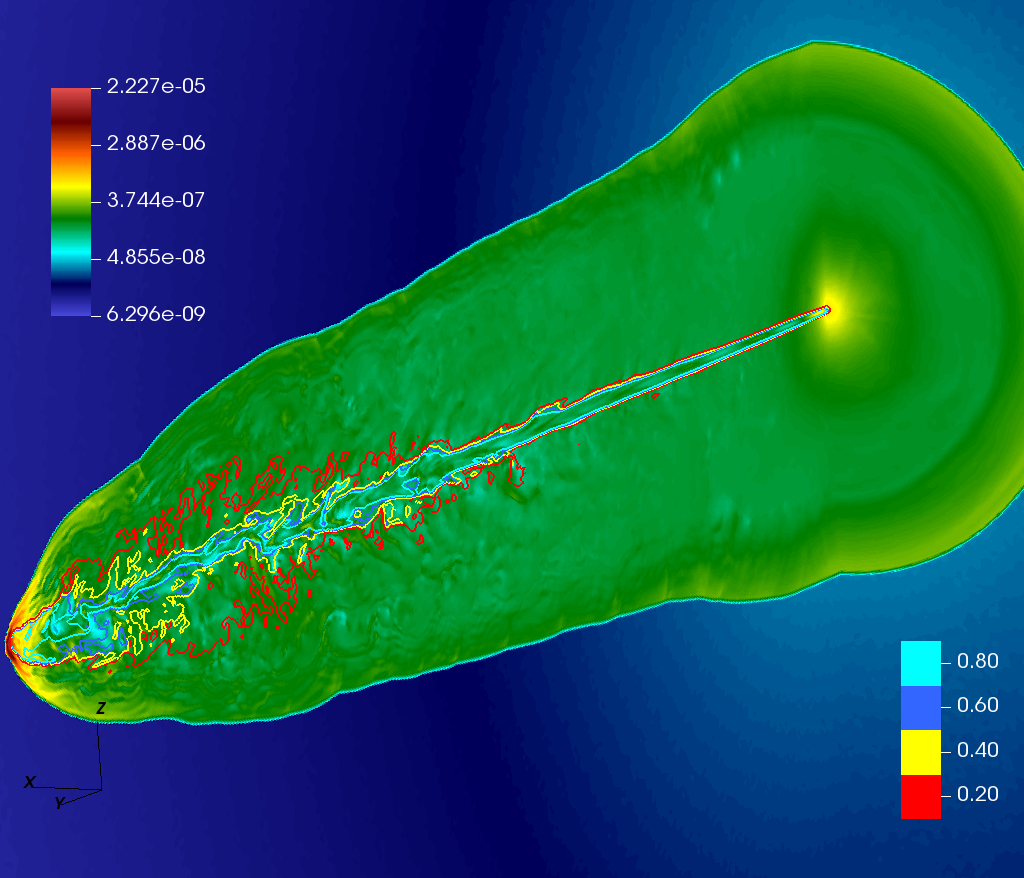}\\
\includegraphics[width=0.45\textwidth]{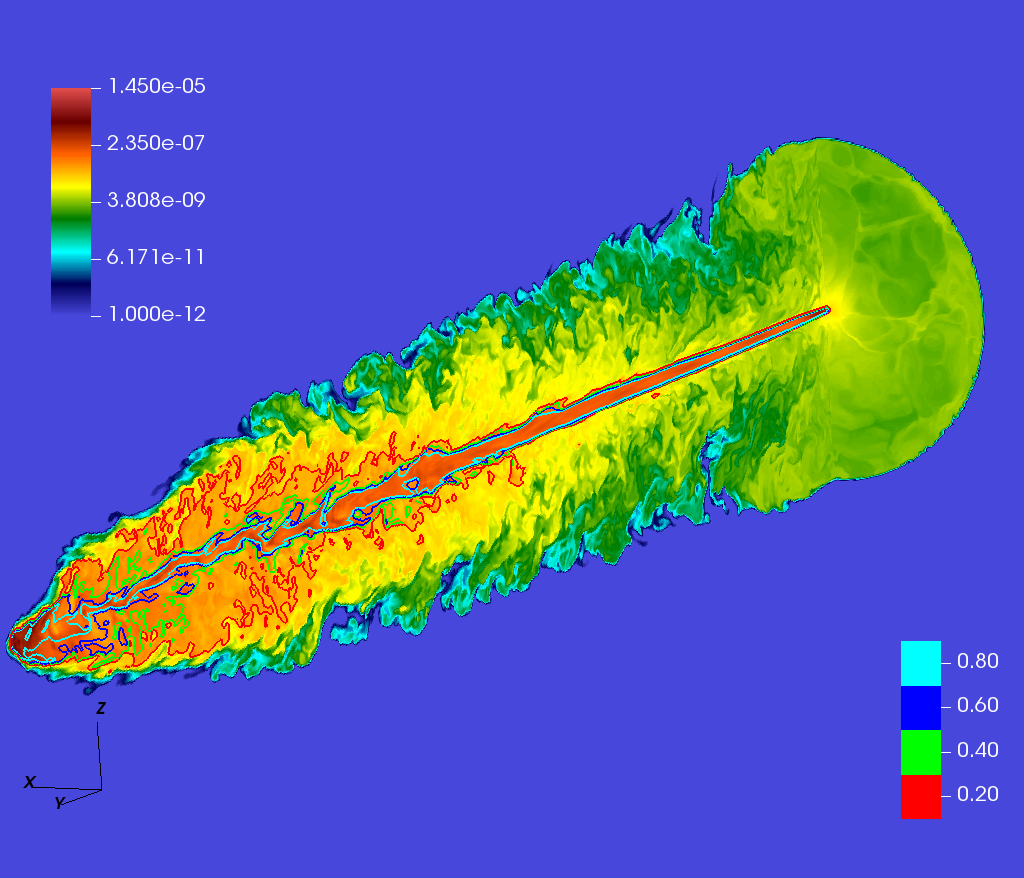}
\caption{From top to bottom, distributions of rest-mass density, pressure, and pressure weighted with the tracer, at the end of the simulation. Contour levels tracing the jet mass fraction are superposed to the maps. Quantities are in code units (based on the jet radius, $R_j$, the speed of light, $c$, and the rest-mass density of the ambient at injection, $\rho_{a,0}$). The jet is $120~{\rm kpc}$ long.} 
\label{fig:rho}
\end{figure}
%

\subsection{The large-scale picture} \label{ss:lsp}

Figure~\ref{fig:rho} shows different cuts of the distributions of rest-mass density, pressure, and pressure weighted with the tracer (the jet mass fraction), at the end of the simulation. The maps also include several tracer contours. The top panel of Figure~\ref{fig:rho} can help to identify the structural ingredients forming the jet. The beam is seen as the collimated central spine in light blue slowly expanding until $\sim 100\,{\rm kpc}$, when it shows signs of helical oscillations with nonlinear amplitude (showing up as intermittent disappearance of the beam channel from the plane of the cut). Surrounding the beam and covering the colour scale from dark blue to light red is the cocoon, formed with beam plasma deflected at the reverse shock (located at the head of the jet). Finally, we find a dense shell (in dark red) of shocked ambient gas between the cocoon and the undisturbed ambient medium. The cocoon undergoes little mixing with the shocked ambient medium, located at the cocoon/shocked ambient medium interface. As a consequence, the region closer to the jet remains very underdense with respect to the shocked ambient. Compared with simulation J0 in Paper~I, the global structure of the outflow generated by model J1 has a thicker cocoon and a smaller axial-to-radial length ratio.

\begin{figure*} 
\includegraphics[width=0.48\textwidth]{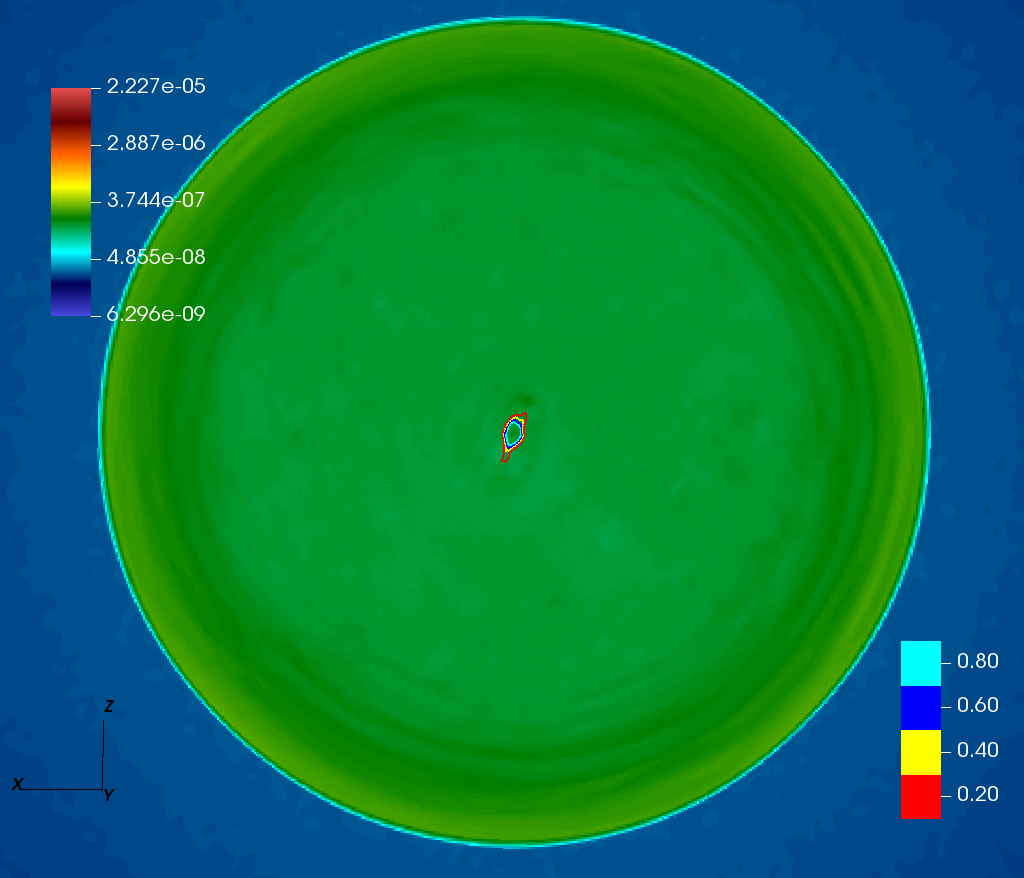}\,\includegraphics[width=0.48\textwidth]{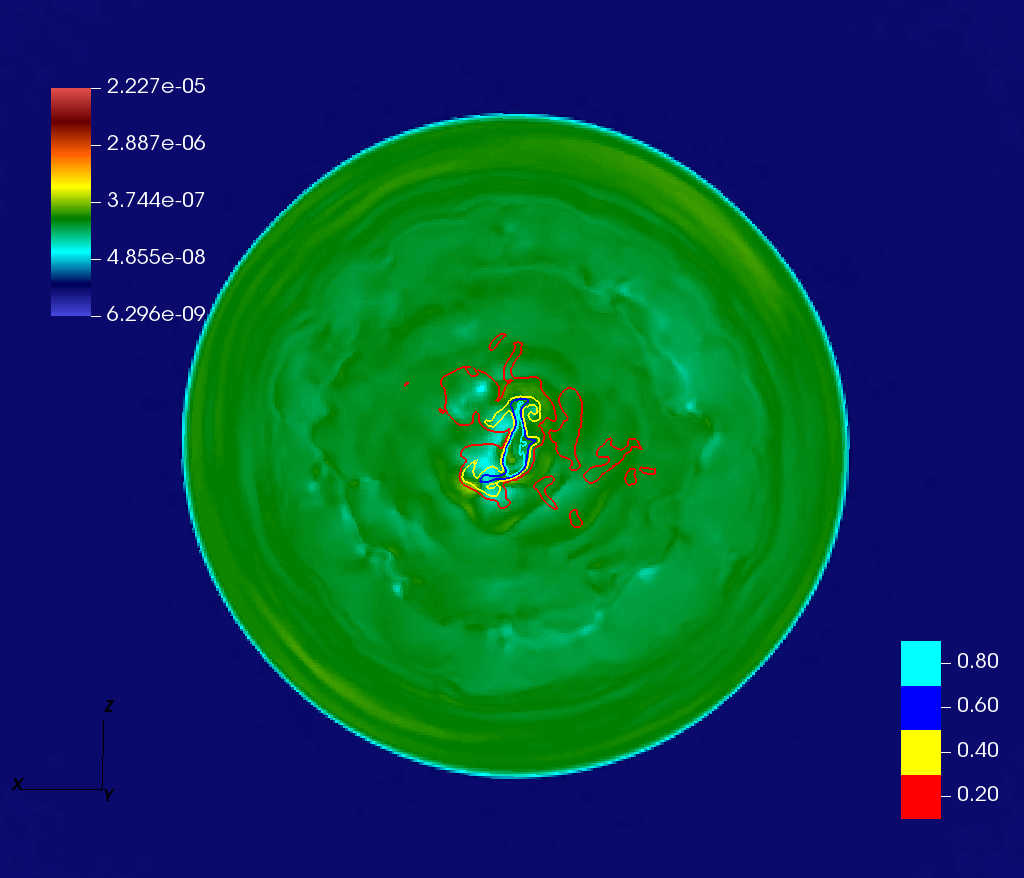}\\
\includegraphics[width=0.48\textwidth]{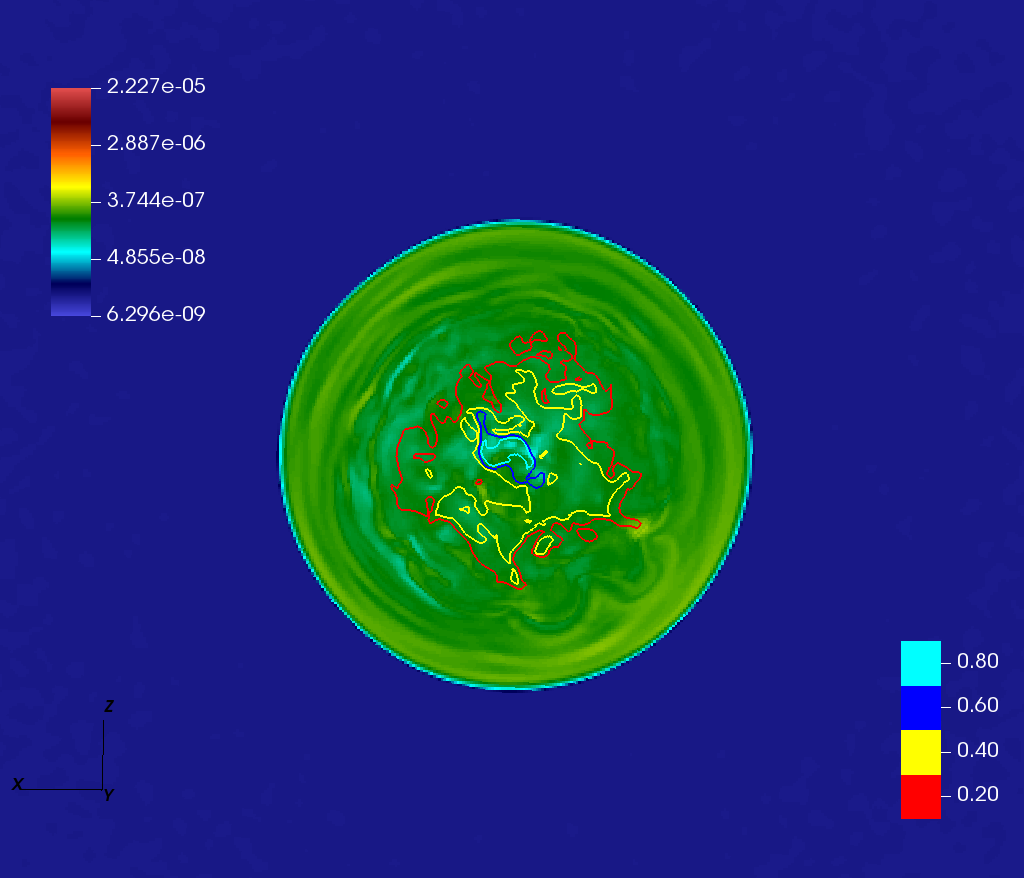}\,\includegraphics[width=0.48\textwidth]{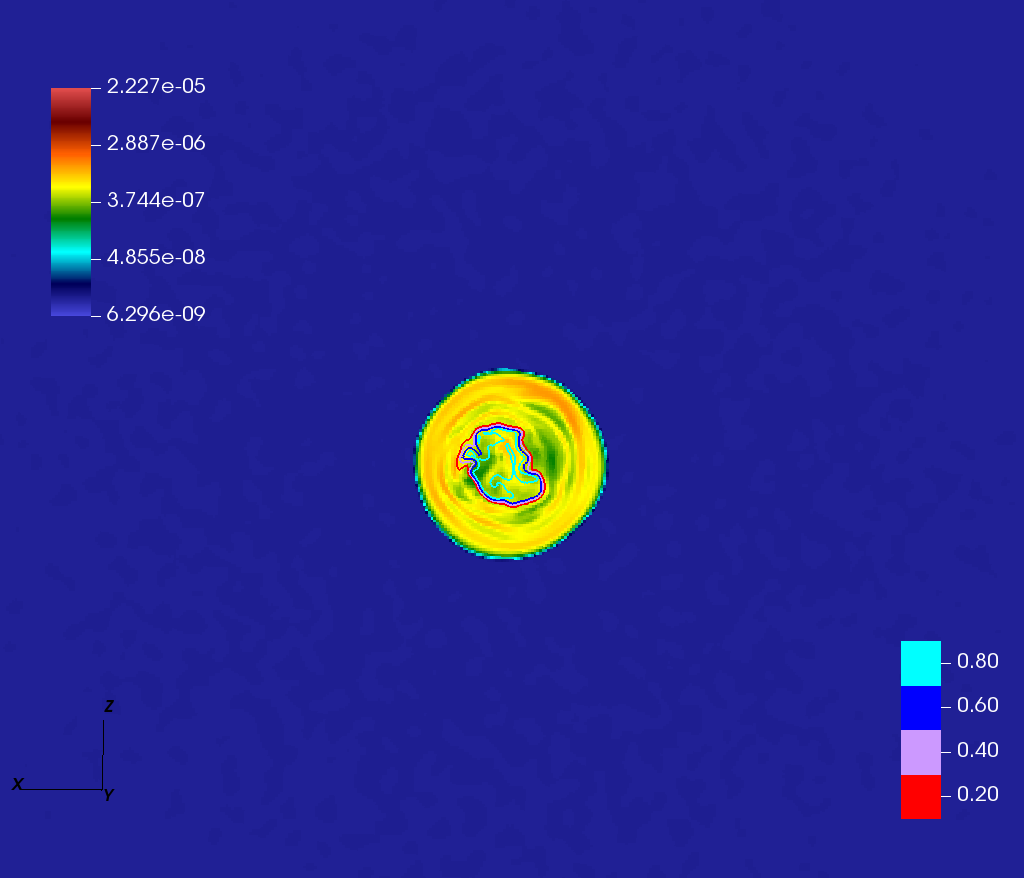}
   \caption{From top to bottom and from left to right, transversal cuts of pressure (code units, $\rho_{a,0} c^2$) and tracer contours at $~25$, $~75$, $~100$ and $~115$~kpc. The figures also display several tracer contours.}  
    \label{fig:pres2}
\end{figure*}

The central panel of Figure~\ref{fig:rho} shows that, as expected from the high sound speeds in the shocked region, the pressure is quite homogeneous but for the terminal shock region. Small pressure enhancements are observed along the jet as darker green regions, caused by expansion and recollimation of the jet. Within the inner 50 kpc, these appear symmetric and elongated, whereas they appear more asymmetric later, revealing the transition from pinching to helical mode dominated jet structure. The transversal cuts displayed in Figure~\ref{fig:pres2} also show the transition from a homogeneous (symmetric) structure to the inhomogeneities produced by the helical oscillations (top right), and to the hotspot at the terminal shock.

\begin{figure*} 
\includegraphics[width=0.45\textwidth]{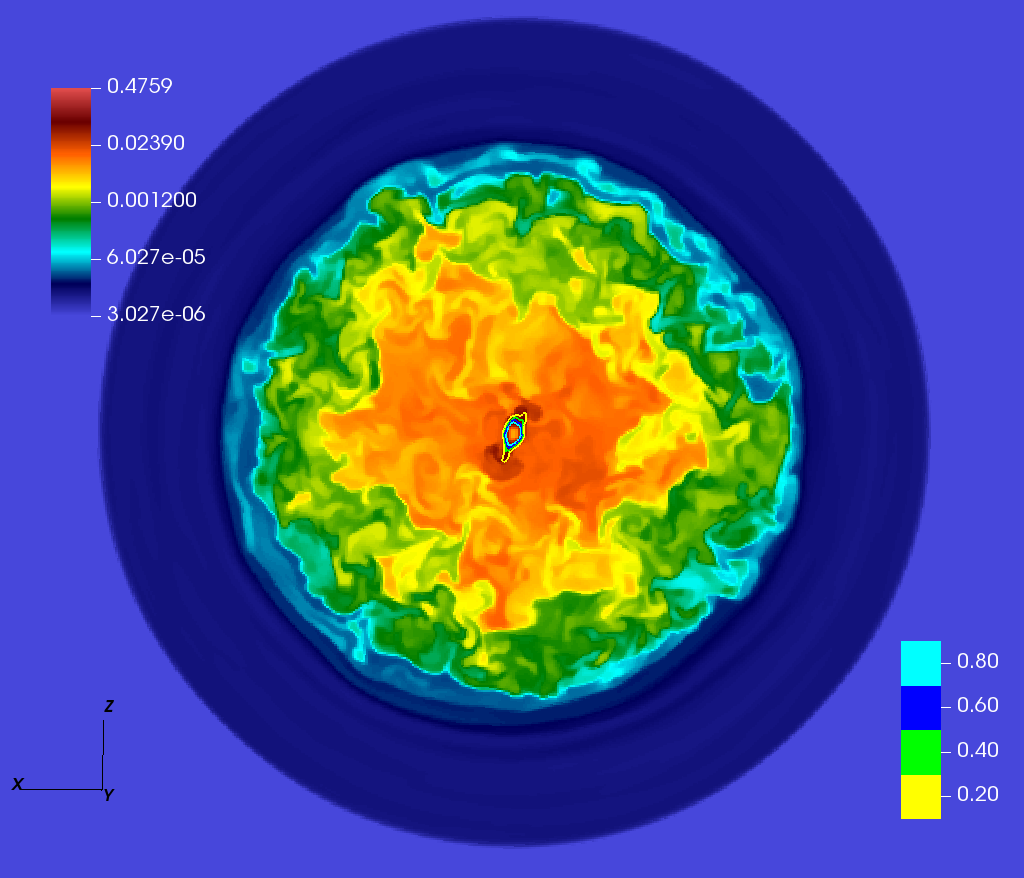}\,
\includegraphics[width=0.45\textwidth]{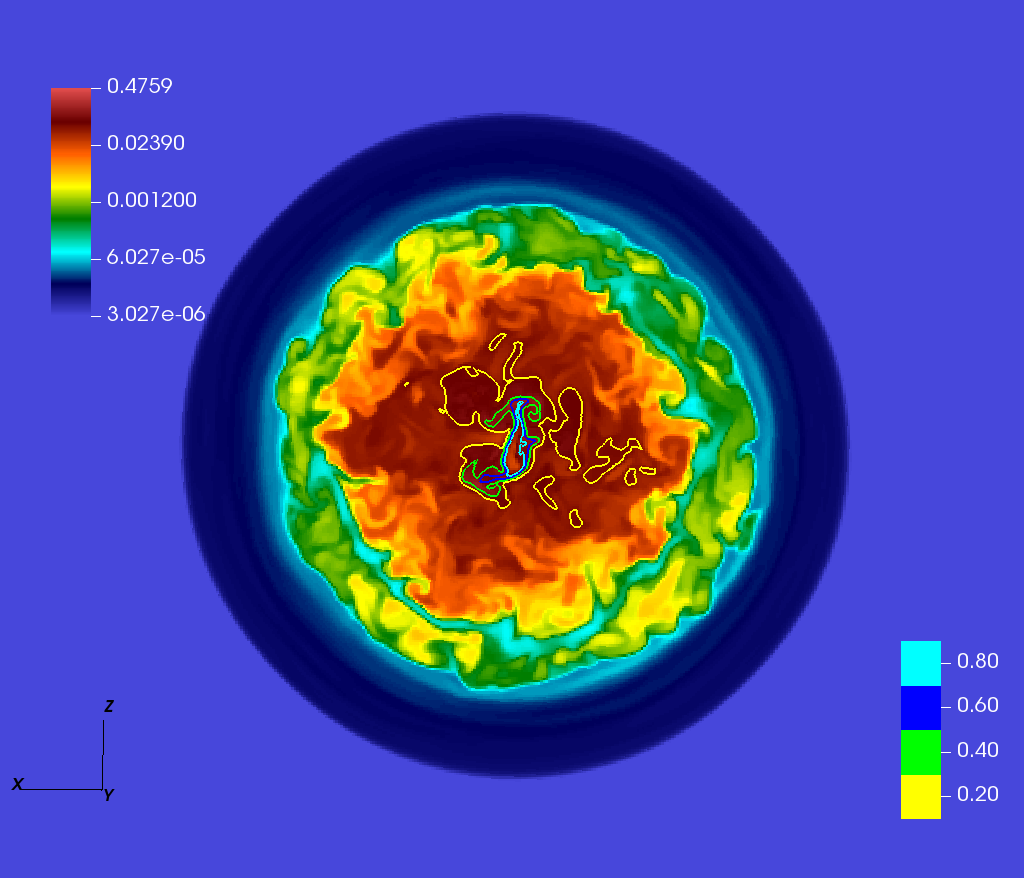}\\
\includegraphics[width=0.45\textwidth]{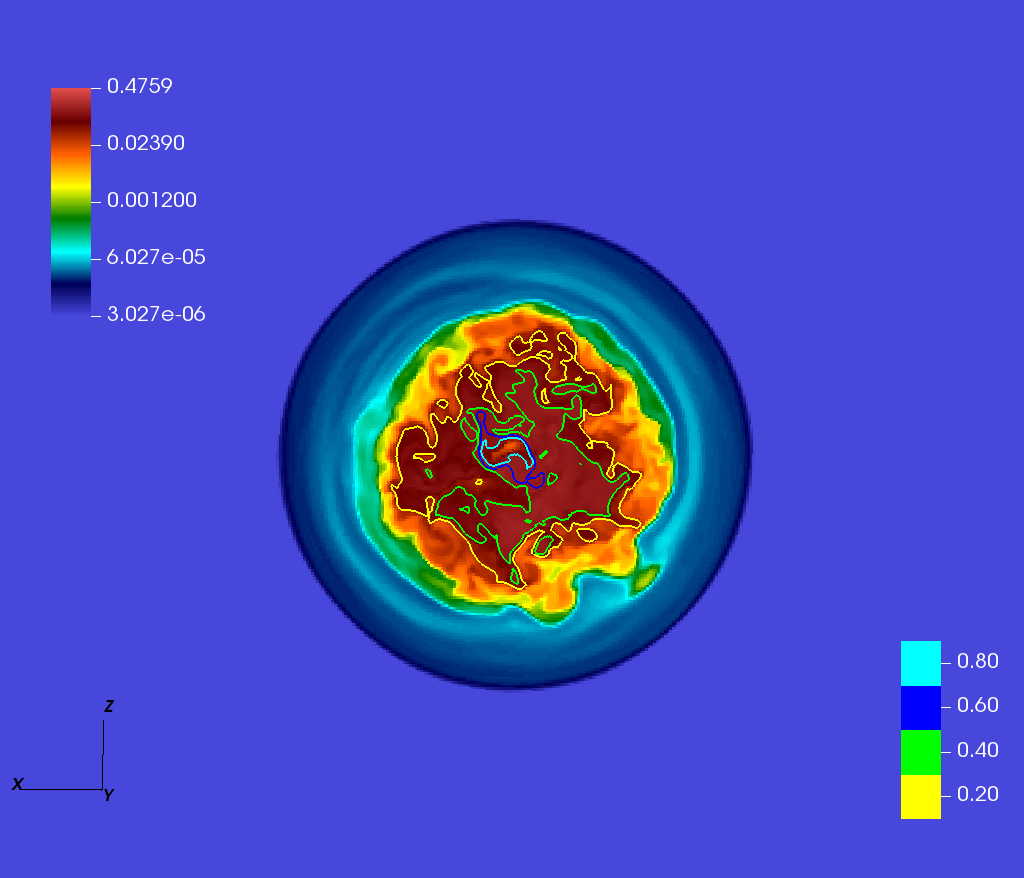}\,
\includegraphics[width=0.45\textwidth]{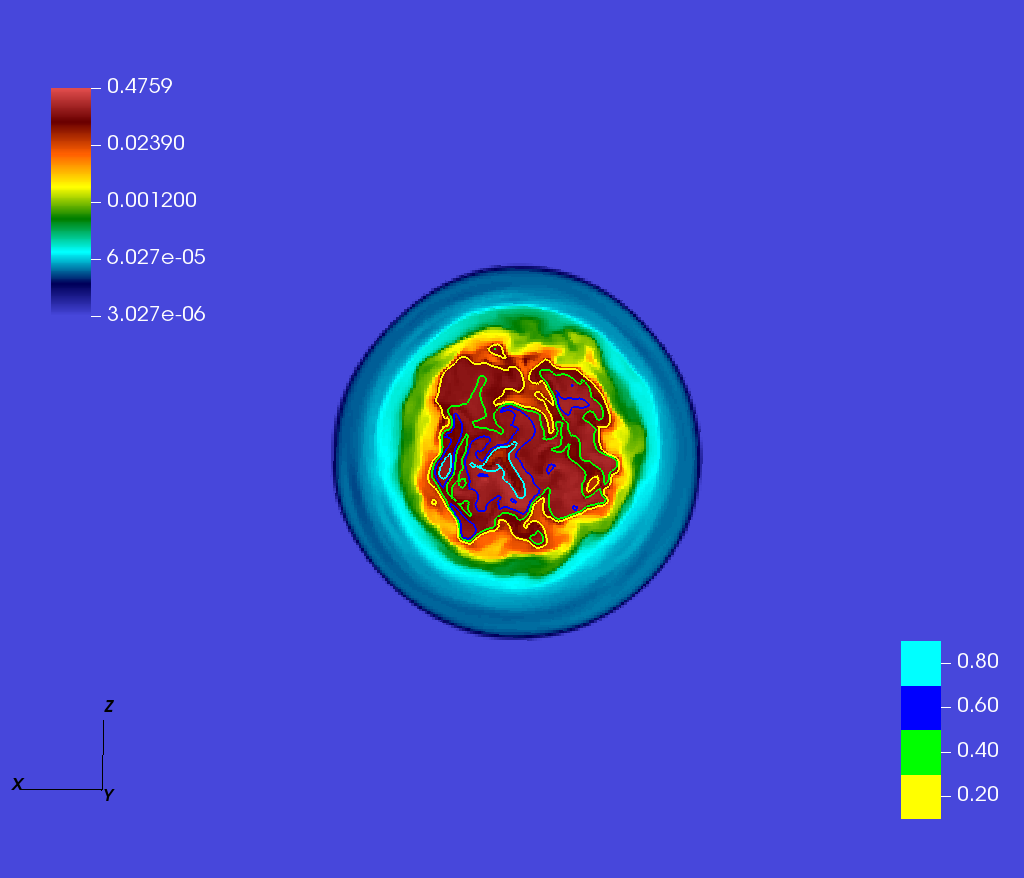}
\caption{From top to bottom and from left to right, transversal cuts of specific internal energy (units of $c^2$) and tracer contours at $~25$, $~75$, $~100$ and $~110$~kpc.}  
\label{fig:eps2}
\end{figure*}

The bottom panel of Figure~\ref{fig:rho} shows the pressure distribution weighted with the tracer. This combination gives the closest approximation to the emitting regions in the system, shown by its highest values. The hottest region, presumably associated to that with higher synchrotron emissivity, is concentrated around the hotspot and within the tracer contours, i.e., the region occupied by the shocked jet plasma. This region is the simulation equivalent to the radio lobes observed at GHz frequencies. At low radio frequencies, as those provided by, e.g., LOFAR, the colder, extended backflow down to the host galaxy ($r\leq 50$ kpc) could be observed \citep{min19}. Figure~\ref{fig:eps2} shows transversal cuts of the specific internal energy, including tracer contours. These contours reveal the sideways expansion of the jet with distance. The colour distributions also show the rise in specific internal energy as the jet head is approached, and that the shocked jet plasma surrounding the jet channel is as hot as, or even hotter, than the jet plasma itself close to the reverse shock. This fact indicates that the energy flux carried by the jet is efficiently dissipated at this shock, in agreement with the expected edge-brightening of FRII radio galaxies. Nevertheless, because the jet is kinetically dominated at injection, we observe that the internal energy is larger at the boundary layer along the jet, where dissipation takes place due to shear and the growth of instabilities in the linear regime \citep{alo99}.

%
%

%
%

\begin{figure*} 
\includegraphics[width=0.8\textwidth]{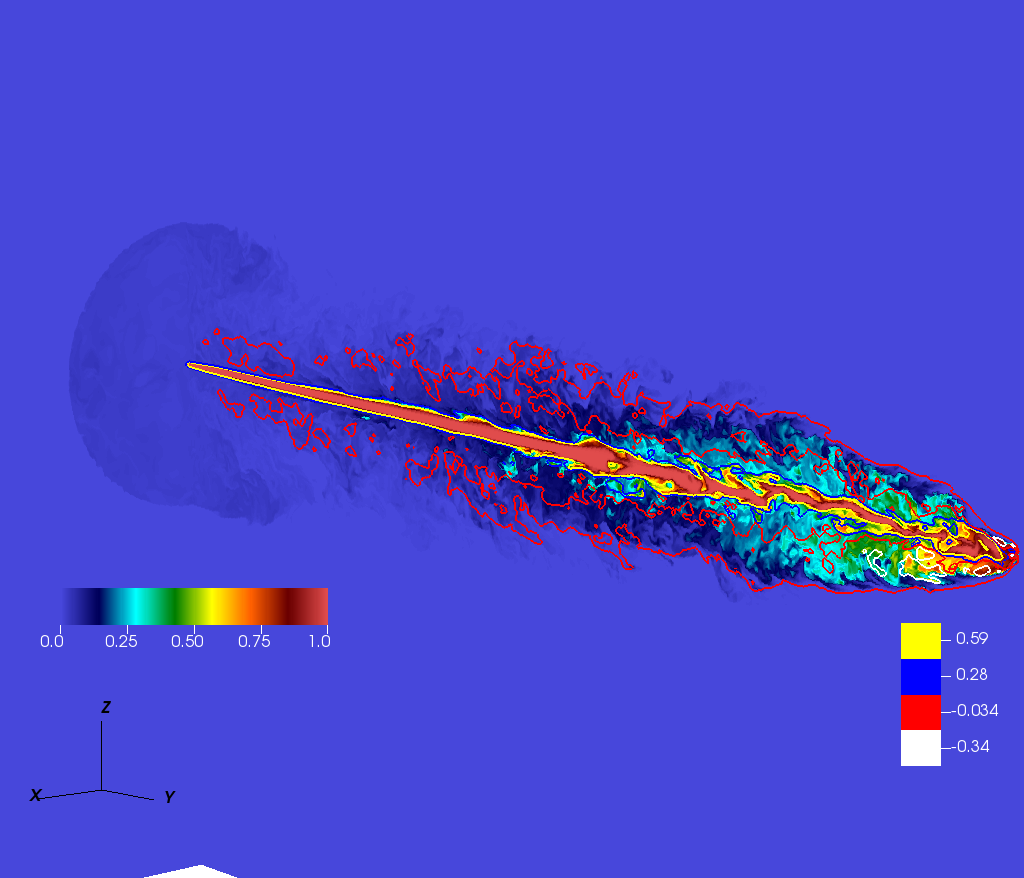}
   \caption{Tracer map plus velocity contour (code units, $c$) at the same frame as Figure~\ref{fig:rho}.}  
    \label{fig:tracer}
\end{figure*}

\begin{figure} 
\includegraphics[width=\columnwidth]{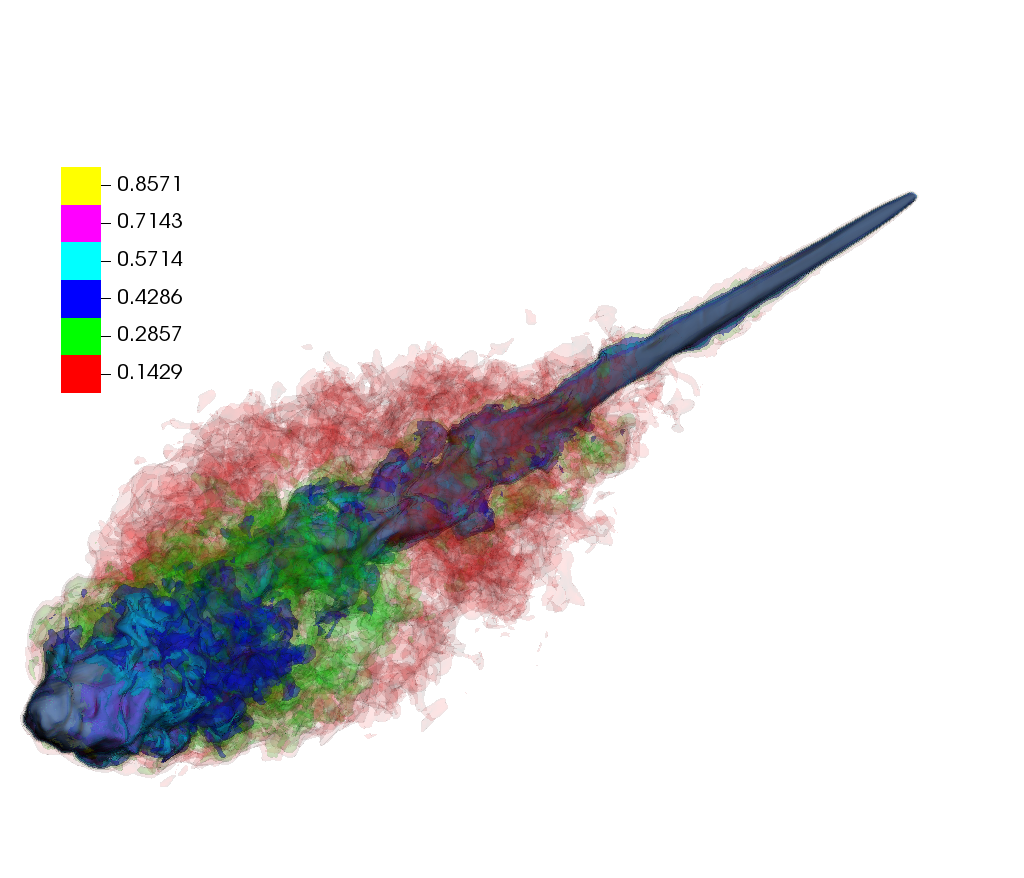}
\caption{Three-dimensional contours of the tracer, showing levels that start at 0.14. A certain degree of transparency is applied to the outer levels to facilitate the visualization of the higher values that trace the jet flow (e.g., $>0.4$).}  
\label{fig:trcont}
\end{figure}

Figure~\ref{fig:tracer} shows the distribution of jet mass fraction and contours of axial velocity. Here we can see that, despite the deviations from the initial jet direction along the $z$ axis, there is a well defined outflow structure, which preserves mildly relativistic velocities along this direction. The backflow generated from the jet terminal shock shows peak values (in absolute value) $\leq 0.4\,c$ in small regions close to the jet head, falling with distance to the interaction site due to dissipation and mixing. Finally, Figure~\ref{fig:trcont} displays different three-dimensional tracer contour levels. We have applied selected transparency degrees to the external values in order to facilitate the visibility of the inner regions. This figure shows how jet collimation is slightly lost by the growth of helical modes, which do not completely disrupt the jet by the end of the simulation. The global morphology has a clear resemblance to the radio lobes observed in classical FRII sources.

%
%

%
%

\section{Discussion and Conclusions} \label{sec:discuss}

\subsection{Jet dynamics}
\label{ss:jd}

As we did in Paper~I with model J0, we consider the evolution of simulation J1 as divided into two phases, an {\it initial} phase (Phase~I) and a {\it multidimensional} one (Phase~II), and use the {\it extended Begelman-Cioffi's} model \citep[eBC;][]{bc89,sch02,pm07,pqm11} to check for the consistency of our interpretation. Details of this analysis as well as a comparison with the model J0 can be found in the Appendix. 

Focusing on Phase~I, the main difference in the evolution of jet J1 with respect to J0 is a remarkably milder acceleration in J1 indicating that the acceleration mechanism based on the wobbling of the terminal shock is not operating in this case. This can be understood on the basis of the increase in the jet/cocoon relativistic density ratio (as happens in model J1 with respect to J0) slowing down the growth of Kelvin-Helmholtz (KH) modes during the linear phase \citep{ph04}, which, as concluded in Paper~I, is the mechanism behind this extra acceleration during the early phase of the jet propagation. Furthermore, taking into account that the growth of the KH modes beyond the linear phase contributes (together with other multidimensional processes) to the triggering of the long-term deceleration of the jet, this result could also explain the longer duration of Phase~I in model J1 (2.8 Myr) as compared with J0 (2.0 Myr).

The evolution along Phase~II is characterized by the deceleration of the jet propagation. The origin of this deceleration should be found in the non-linear growth of the KH modes and the triggering of the {\it dentist-drill effect}. Comparison with model J0 shows that the deceleration is milder in model J1. This result is in contrast with the fact that KH modes are more disruptive for larger values of the jet/cocoon density ratio (in our case, in model J1 with respect to J0). However, the studies of the growth of KH instabilities are done for flows in pressure equilibrium with the ambient. In contrast, in our simulations, jets are immersed in overpressured cocoons, which can inhibit the growth of unstable modes. The fact that the overpressure factor between the cocoon and the jet in model J1 (see Figure~\ref{fig:pres2}) is larger than in model J0 could explain a slower development of the disruptive KH modes in model J1 and the subsequent slower deceleration.

From a thermodynamical point of view, both the increased ISM/IGM pressure and the slower expansion cause an increase in the cocoon density and pressure. A direct consequence of this would be that lobes are much brighter (as it is the case in e.g., Cygnus~A).

We stress here that, despite the increased interaction intensity in J1 with respect to J0, the efficiency of the energy transfer to the ambient medium remains similarly high. This implies that the efficiency is related to jet properties, and probably related to its relativistic nature and collimation, as anticipated by \cite{pmqb17}. Following this work, jets with large opening angles or with large overpressure with respect to the ambient medium would show smaller efficiencies of shock heating with time, but this is not the situation in powerful extragalactic outflows, even in young FRI jets \citep{pm07}.

\begin{figure*} 
\includegraphics[width=0.45\textwidth]{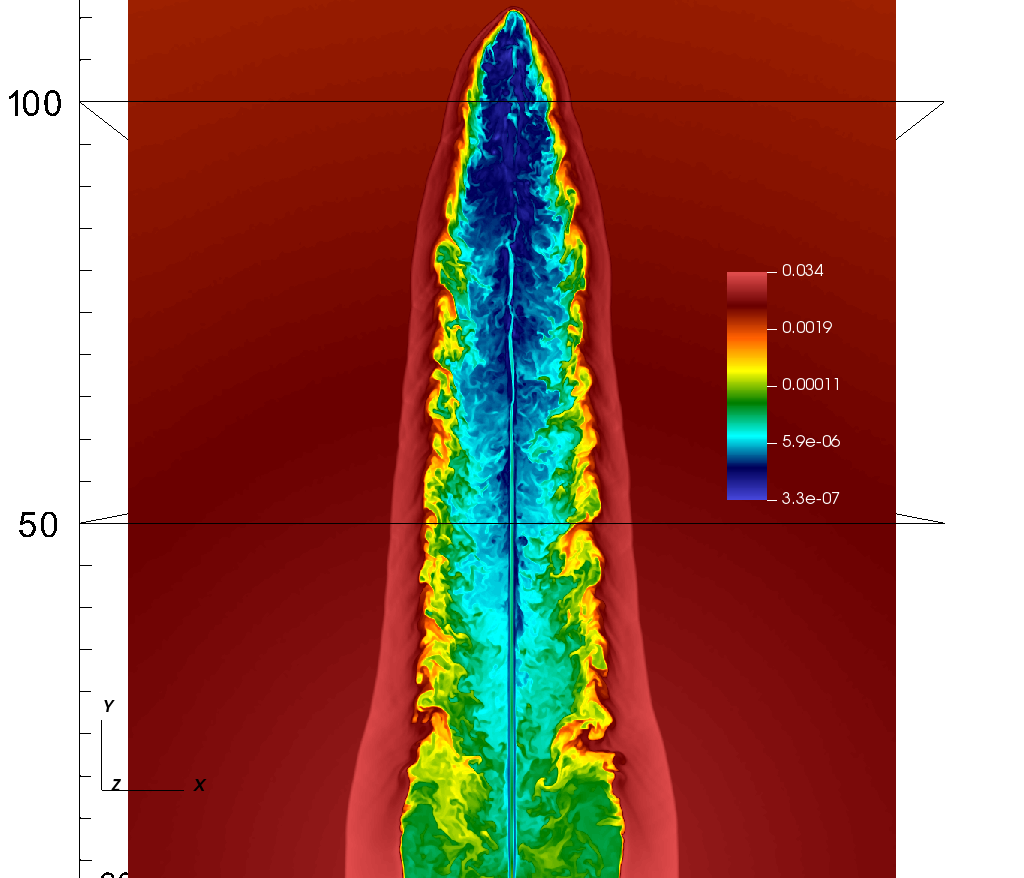}\,
\includegraphics[width=0.45\textwidth]{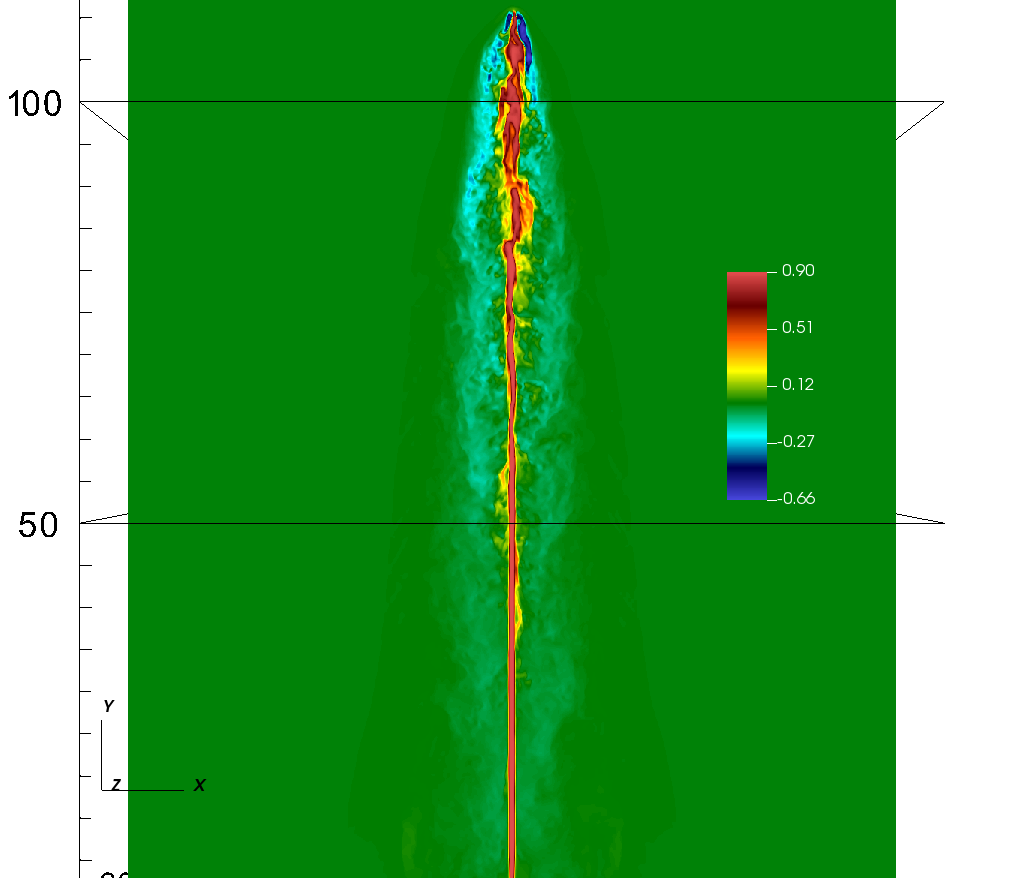}
\caption{Cuts in logarithm of rest-mass density (left), and axial velocity (right) in code units. The axial distance is in kpc.}  
\label{fig:cuts}
\end{figure*}

Figure~\ref{fig:cuts} shows cuts in rest-mass density and axial velocity at the end of the numerical simulation ($t = 8$ Myr) revealing the development of helical instabilities that become nonlinear beyond $\sim 60~{\rm kpc}$ in axial distance. The observed transition to non-linearity is the evolution of the original instability that caused the onset of the deceleration of the jet at $t= 2.8$ Myr. The growing helical structure seen in the Figure has a wavelength that is much larger than the jet radius.

%
%

In order to study the stability properties of both jets and their different evolution, we have solved the differential equation for the development of linear perturbations in the system (see Appendix~\ref{a:sa}). The left panels of Figure~\ref{fig:KH} show the radial profiles of specific internal energy and rest-mass density used as inputs for the numerical solver of the stability equation (note the presence of a hot shear layer in the transition between the jet and the cocoon media in both models). The axial velocity, $\sim 0.9\,c$, and the corresponding Lorentz factor, also required by the equation, have a regular monotonic fall across the shear layer and are not shown. These profiles are given by functions that approximately reproduce those observed for models J0 and J1 after the initial 10~kpc. At this distance, the jet is already well established but the growing instabilities have not still developed to the non-linear regime, so we can use those profiles to explore the properties of the linear perturbations that can grow in the system\footnote{In a dynamical simulation, the radial profiles of the physical quantities change slightly in space and time, even close to the jet base. The profiles shown in Figure~\ref{fig:KH} are simple approximations of the averaged profiles for a number of radial directions and several axial distances around 10 kpc taken at some particular time. On the other hand, these small quantitative changes in the profiles of the physical parameters do not affect the overall stability of the system \citep[see, e.g.,][where perturbation modes are tracked for flows with large changes in the physical parameters and profiles]{vg19}. The crucial point in the analysis of the stability of the two models is the difference in the ambient density (captured in the approximated density profiles of Figure~\ref{fig:KH} outside the jet).}. The cocoon pressure has been taken as the equilibrium pressure of the system, as required by this analysis (see Appendix~\ref{a:sa}). 

The equation that we solve is derived under the assumption of an infinite flow in the axial direction. Therefore, we have chosen the temporal approach (the modes are only allowed to grow in time) because in our simulations the spatial development of instabilities (the growth in distance) is conditioned by the jet head propagation, which is different for both jets. 

The results to the stability problem for both jets are shown in the right panels of Figure~\ref{fig:KH} where the real ($w_r$, upper curves) and imaginary ($w_i$, lower curves) parts of the unstable frequency are plotted against the mode wave-number, $k$. In the temporal approach to the stability problem, the imaginary part of the frequency represents the inverse of the $e$-folding time for the amplitude of the unstable wave, so the larger the values of $w_i$, the faster the growth of the unstable modes in time. As anticipated, the growth times are longer (the values of $w_i$ typically smaller) for J1 at the relevant long wavelengths -small $k$-, which means that this jet takes longer to develop instabilities. The maximum growth rates at the longest, disrupting, wavelengths are given for the surface modes, with $w_{i,{\rm max}}\simeq 0.02\,c/R_j$ ($\sim 6.1\times 10^{-6}\,{\rm yr^{-1}}$) for J1, and $\simeq 0.03\,c/R_j$ ($\sim 9.1\times 10^{-6}\,{\rm yr^{-1}}$) for J0. The factor $\sim 1.5$ between the linear growth rates of the two models can explain the differences found in the amplitudes of the instabilities in the long term. 

The stability problem has also been solved for the standard, vortex-sheet case (not shown in the figures), i.e., without the presence of shear layers and no hot region at the jet boundaries. The comparison of the solutions for the two different radial distributions of physical parameters shows that the presence of the hot shear-layer contributes to the long-term stability of the jet by increasing the growth-times of all unstable modes, mainly in the case of short-wavelength, body modes. However, the growth of modes with long wavelengths ($k \ll 1/R_j$) -the fastest growing modes seen in our simulations (see Figure~\ref{fig:KH})-, are not affected by the presence of such shear-layers. These hot shear layers can be naturally produced by viscous shear dissipation, or also by the growth of small scale KH resonant modes at this transition, as shown by \citet{pe07}, and have radiative implications \citep[see, e.g.,][and references therein]{rl18,rd19}.

%
\begin{figure*}
\includegraphics[trim=3cm 13cm 0cm 4cm,width=\columnwidth]{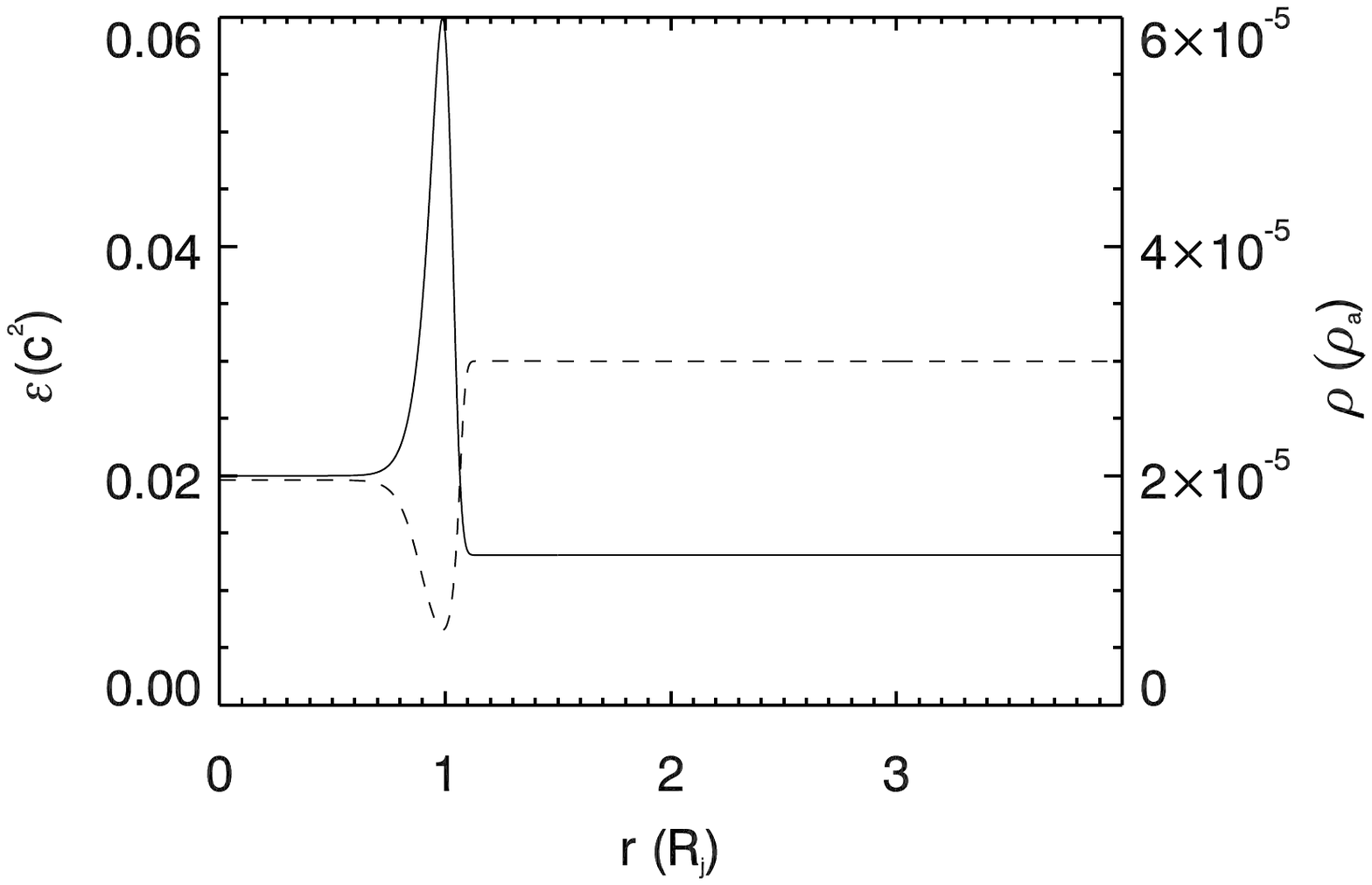}\,
\includegraphics[trim=3cm 13cm 0cm 4cm,width=\columnwidth]{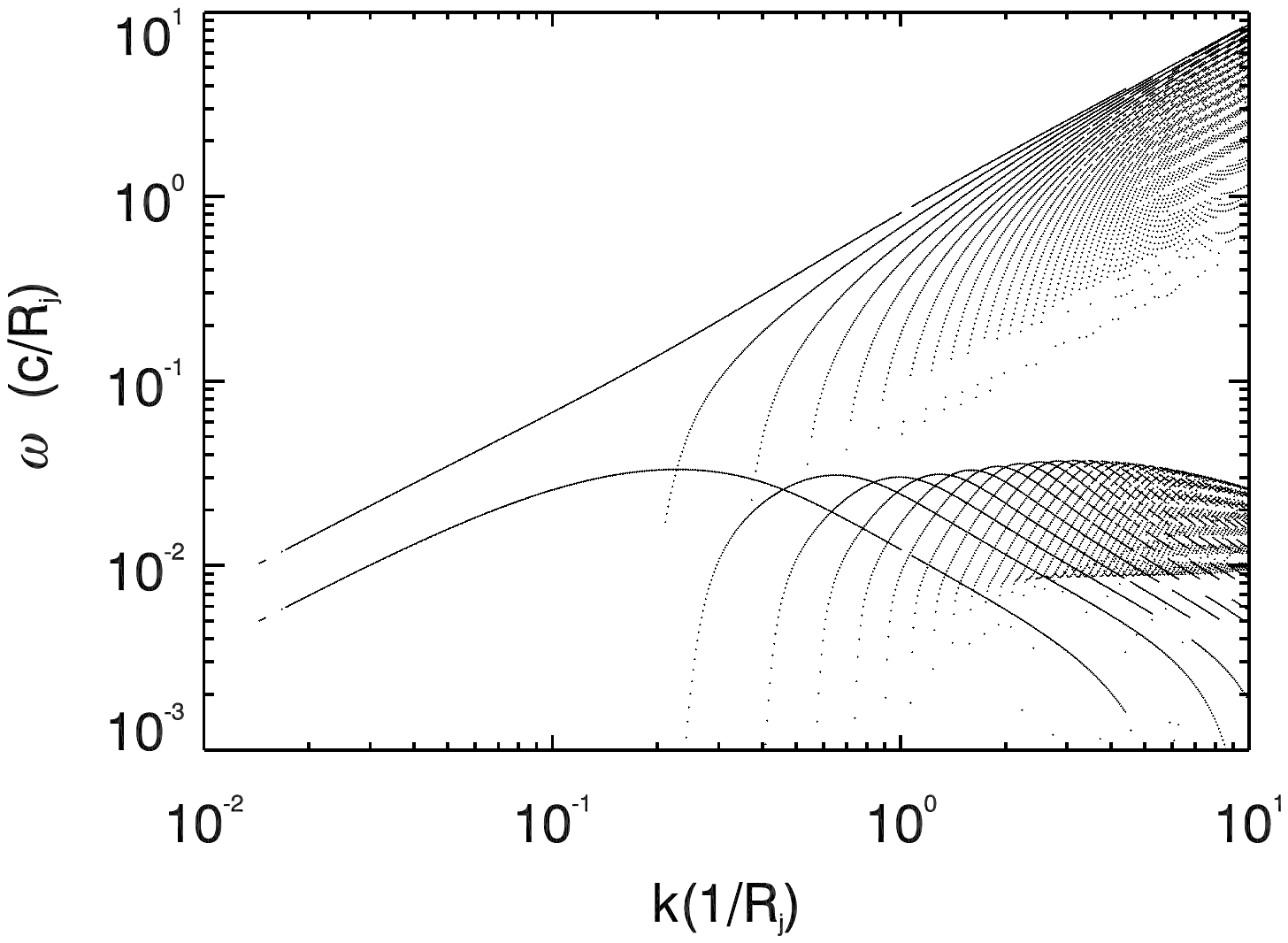}\\
\includegraphics[trim=3cm 13cm 0cm 4cm,width=\columnwidth]{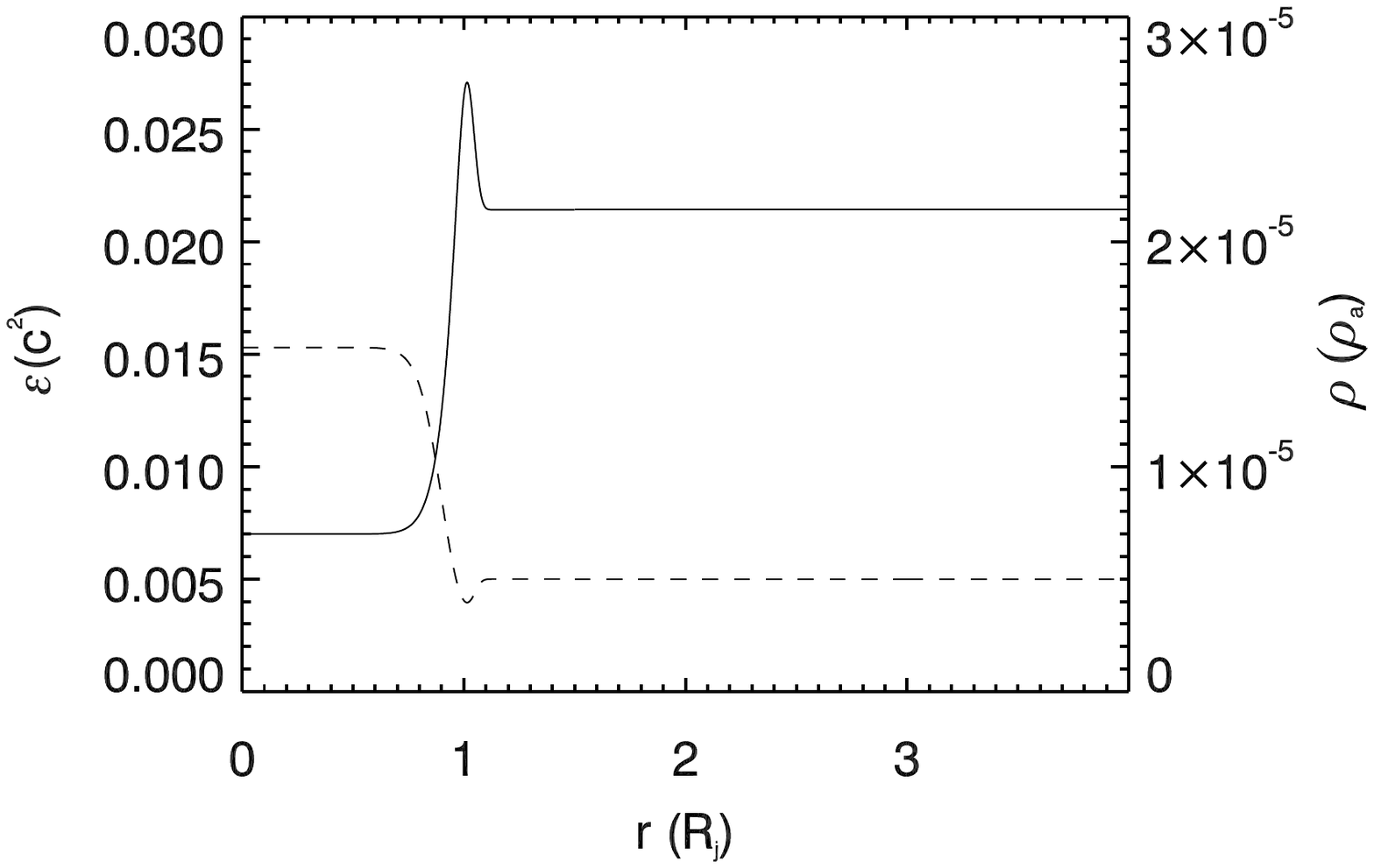}\,
\includegraphics[trim=3cm 13cm 0cm 4cm,width=\columnwidth]{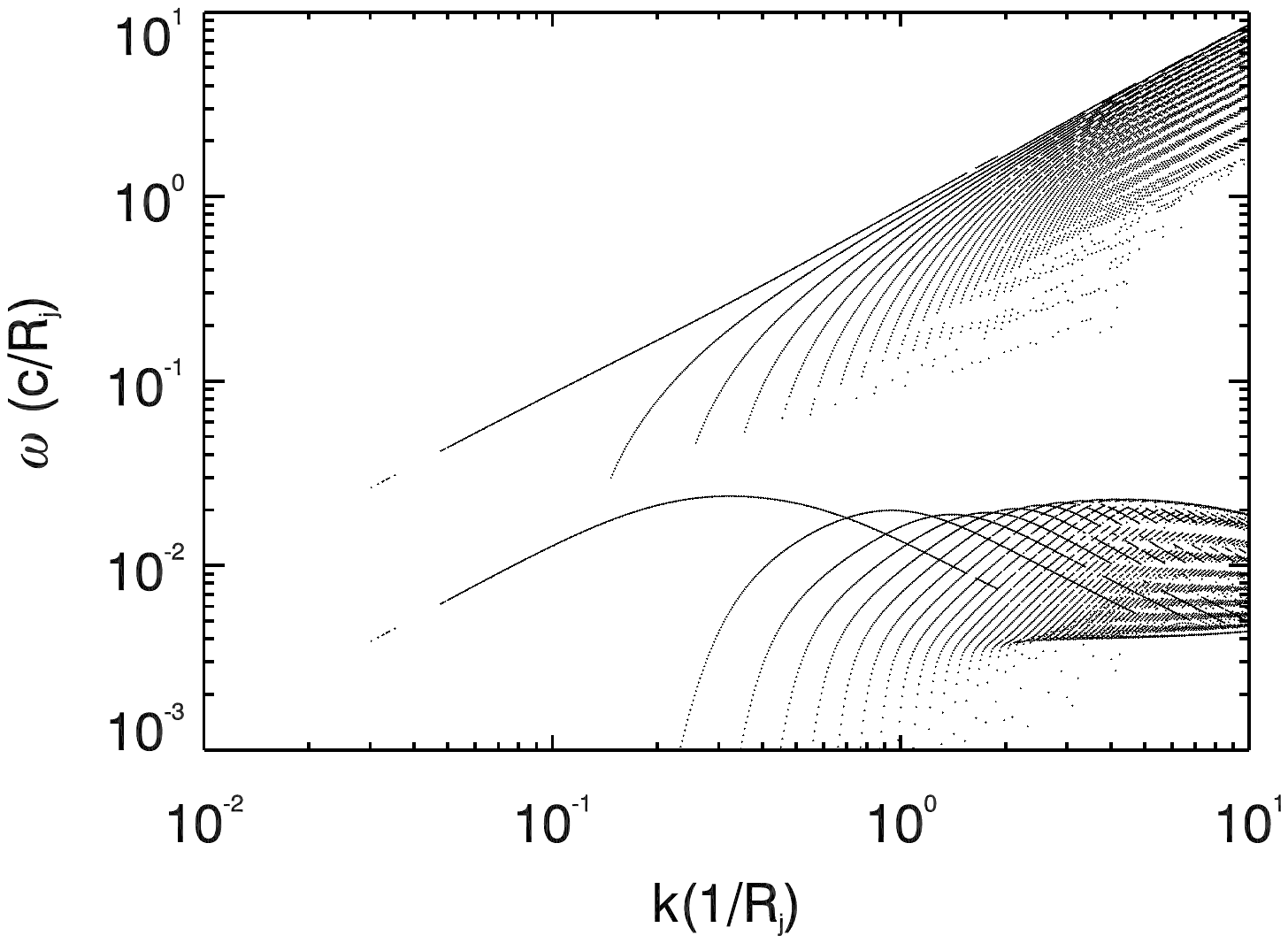}
\caption{The left panels show approximated specific internal energy (solid line) and rest-mass density (dashed line) radial distributions for the jets J0 (top left) and J1 (bottom left). The right panels show the distribution of unstable helical modes in the $k-\omega$ diagram, with the upper curves in each panel standing for $w_r$ (the real part of the frequency), and the lower ones for $w_i$ (the imaginary part of the frequency). The numerical solver \citep{pe07,vg19} finds the roots in a point-to-point way. This results in some of the points possibly being missed and the discontinuous appearance of the curves for each mode, which should be continuous otherwise.}
\label{fig:KH}
\end{figure*}
%

\begin{figure*} 
\includegraphics[width=0.45\textwidth]{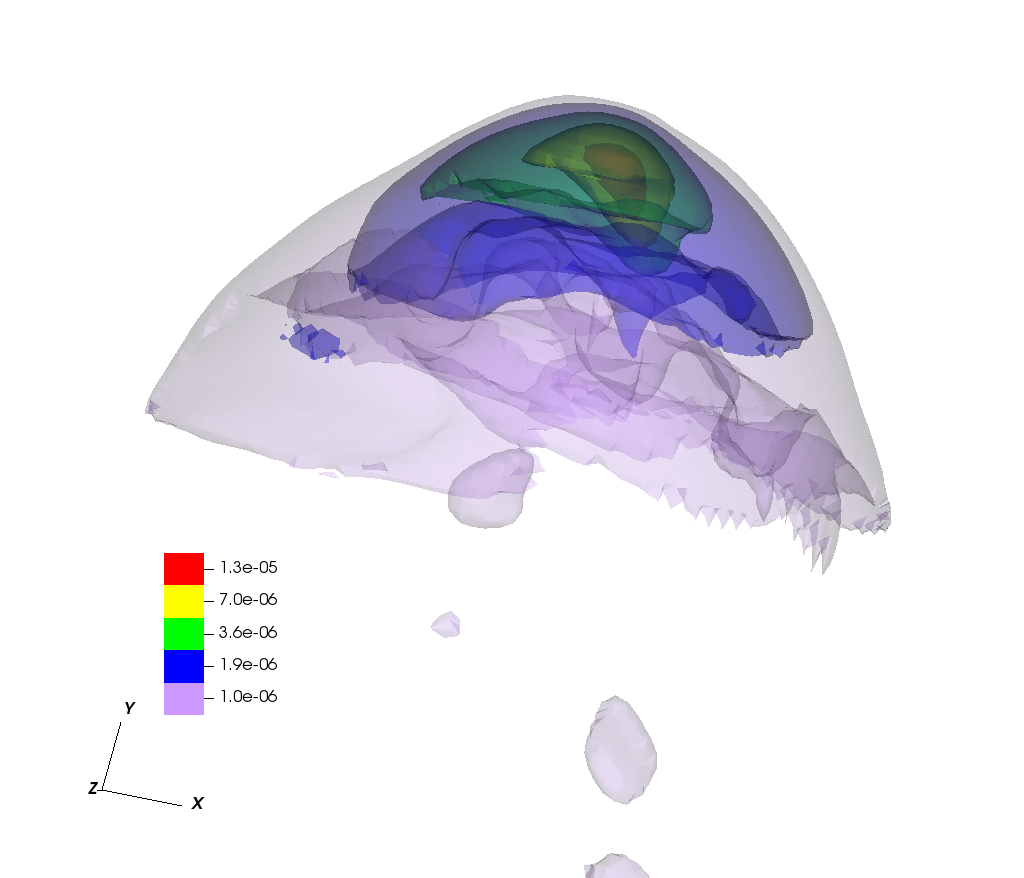}\,
\includegraphics[width=0.45\textwidth]{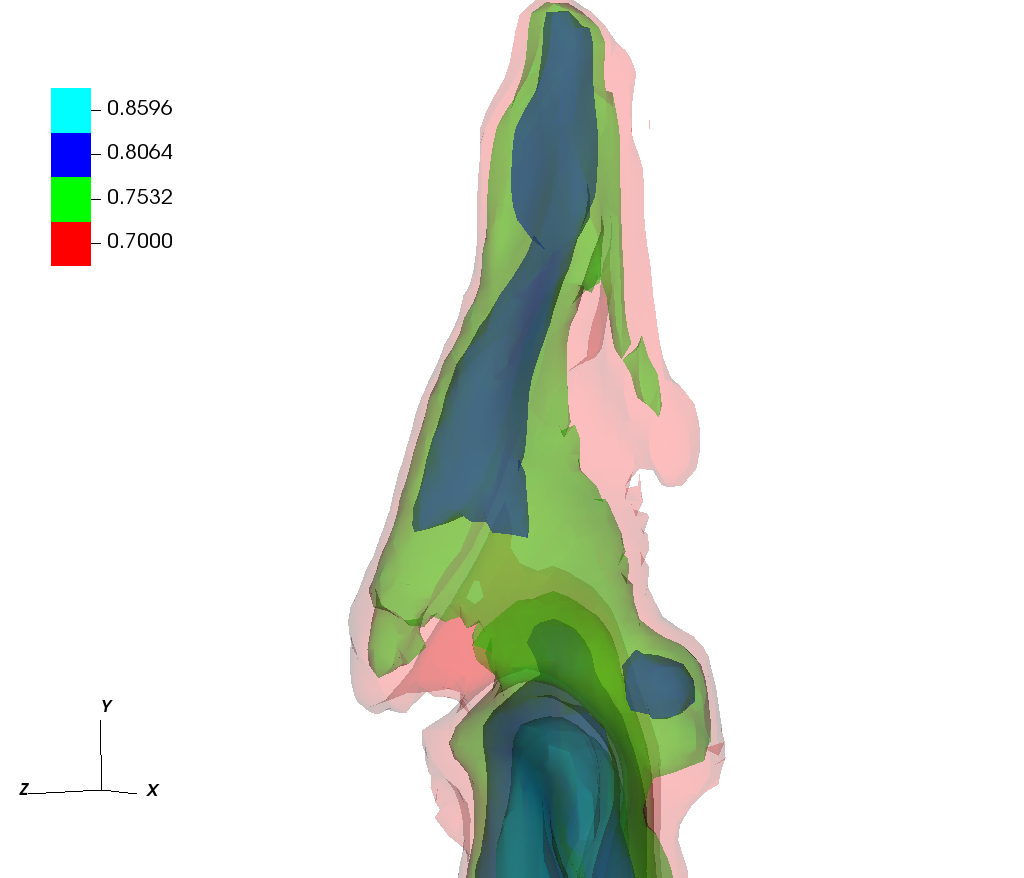}
\caption{Pressure and axial velocity isosurfaces around the jet head. The values are in code units. The viewing angle of each plot has been chosen to enhance the inhomogeneities in the corresponding quantity.}  
\label{fig:hotspot}
\end{figure*}

Figure~\ref{fig:hotspot} shows the pressure and axial velocity isosurfaces around the jet head at the last snapshot of the simulation. The images include, approximately, the last 10~kpc of the jet before interaction with the ambient medium. Obviously, the highest pressure is found at the point where the flow impacts, but it is surrounded by a wide and complex region with high pressure values. The complexity of the region is also shown by the velocity isosurfaces, with detached regions with high velocities, which trigger internal shocks. This kind of extended, complex structure has also been reported and studied by other authors \citep{tre04,ma19,mu21} in simulations limited to smaller grids and/or physical sizes. The hotspot structure extends across a region with a diameter $\simeq 5\,{\rm kpc}$, in agreement with hotspot sizes reported in \citet{js00,pm03,kk06,mig20,mk16}, and it has a fair morphological resemblance to detailed hotspot radio images as those shown in, e.g., \citet{mig20} or \citet{mk16}.

%
%

\subsection{Comparison with FRII sources from the 3C Catalogue}\label{sec:frii}

In Table~\ref{tab:tab3}, we list a series of parameters that can be derived from observations and which can thus be compared to FRII sources (see Table~2 in Paper~I). The Mach number listed in the last row was obtained by means of Equation~15 in Paper~I \citep[from][]{wb06}. The increased ambient density and pressure allows us to compare the results from simulations J1 with jets as those in 3C~228, 3C~244.1, 3C~267 or 3C~427.1 \citep{od09,sni18}. These sources have $L_{\rm BS}\simeq 100 - 200\,{\rm kpc})$, $\rho_{\rm ICM}\sim 10^{-27}\,{\rm g/cm^3}$, $P_{\rm ICM}\sim 10^{-11}\,{\rm dyn/cm^2}$, $T_{\rm ICM}\simeq 2 - 3 \, {\rm keV}$,  $P_{\rm l}\simeq 10-100\times 10^{-11}\,\,{\rm dyn/cm^2}$ ($3.6\times 10^{-11}\,\,{\rm dyn/cm^2}$ in the case of 3C~228, 3C~244.1), $V_{\rm l}\simeq 3-4\times 10^5\,{\rm kpc^3}$ in 3C~244.1 and 3C~427.1 ($55\times 10^5\,{\rm kpc^3}$ for 3C~228), $v_{\rm hs}\sim \rm{10^{-2} c}$ for 3C~427.1 ($\sim 3\times\rm{10^{-4} c}$ for 3C~267), and $M \simeq 1.1 - 2.7$. These values are closer to those shown in Table~\ref{tab:tab3} for J1 than to those for J0, and typically within the same decade of each parameter. The ambient pressure and density of our simulations and the sources used for comparison still lie an order of magnitude below those estimated for the FRII radio galaxy Cygnus~A, which shows remarkably inflated lobes, relative to other FRIIs. This trend again shows the direct relation between the environmental properties and jet lobe shapes, which can be used as a probe of ICM around radio galaxies and quasars. Actually, although the estimated properties of the ICM media in those radio galaxies are similar to those in our simulations, the estimated ages in observational modelling are longer than the time required by J1 jets to reach 100~kpc. This can be due, on the one hand, to an over-estimate of ages from spectral ageing, although particle reacceleration at shocks suggests that these estimates represent lower limits. On the other hand, the imposed collimation of the simulated jets at injection favours rapid expansion. The inclusion of an opening angle in the jet at injection would certainly force a drop in the advance speed of the flow, as discussed in \cite{pmqr14} and in \cite{pmqb17}, thus increasing the time needed by the jet to reach those scales and approaching the ages obtained from both methods.

%
\begin{table}
  \begin{center}
 {\small
  \begin{tabular}{l | ccc}\hline
  & J0 & J1 & J1 \\
  \hline
$T\,[{\rm Myr}]$& 5.4 & 5.4 & 8 \\
$L_{\rm BS}\,[{\rm kpc}]$& 200 & 90 & 118 \\
$\rho_{\rm ICM}\,[10^{-28}\,{\rm g/cm^3}]$ & 4 & 36 & 20\\
$P_{\rm ICM}\, [10^{-12}\,{\rm dyn/cm^2}]$ &1 & 9 & 5 \\
$T_{\rm ICM}\,[{\rm keV}]$ &1.5& 1.5 & 1.5\\
$P_{\rm l}\, [10^{-11}\,{\rm dyn/cm^2}]$ &  5 & 16 & 10  \\
$V_{\rm l}\,[10^5\,{\rm kpc^3}]$ & 0.4 & 0.1 & 0.23 \\ 
$v_{\rm hs}\,[\rm{10^{-2} c}]$ & 12 & 5 & 3 \\
$M$ & 6.3 & 3.8 & 4.0 \\
\hline
  \end{tabular}
  }
   \caption{Cocoon parameters in J0 (Paper~I) and J1 (this paper) at two different times -end of simulation J0 and end of simulation J1. 1st row: Jet age. 2nd row: Linear size. 3rd row: ICM density close to $L_{\rm BS}$. 4th row: ICM pressure close to $L_{\rm BS}$. 5th row: ICM temperature. 6th row: Cocoon (lobe) pressure. 7th row: Cocoon (lobe) volume. 8th row: advance velocity. 9th row: Mach number computed as in \citet{wb06}.}
 \label{tab:tab3}
 \end{center}
\end{table}
%

\subsection{Conclusions}

The results that we have obtained in Paper~I and here point towards radio-lobe morphologies and temporal jet evolution in FRIIs (e.g., in terms of the development of jet acceleration as the jet transits from the ISM to the IGM) to be strongly dependent on the properties of the ambient medium and the development of KH instabilities in the flow. Despite the limited number of simulations that have been run following relativistic jets from the outskirts of the host galaxies to hundreds of kiloparsecs, these have already shed light on the relevant role of the jet-ambient relative properties on the evolution of the radio galaxy and its exact morphology. This means that a fine tune of parameters could provide good qualitative approximations to FRII jets. In particular, the long-term simulation presented here (one of the longest simulations of FRII jet evolution run so far) shows a remarkable morphological resemblance to canonical FRII jets at scales of tens to hundreds of kiloparsecs, including the size and morphology of the hotspot structure.

Future research should focus on other aspects of jet evolution, such as the dissipative processes, the inclusion of magnetic fields, which even if they may be almost negligible from a dynamical perspective play an obvious role in the radiative and dissipative processes at shocks and turbulent regions \citep{mi10,ma19,pe19,mu20}, or numerical experiments of different aspects of jet physics, including microscopic processes \citep[see, e.g.,][]{Nishikawa2021}. 

In constrast, the large-scale morphologies of FRI jets are still difficult to recover from numerical simulations due to the long time-scales involved in their evolution, because they are slow and decollimated, so their expansion is much slower \citep[see][for a simulation of the initial evolution of a FRI jet]{pm07}. The processes involved in jet deceleration and the energy dissipation at FRI jets should also be tackled in future work \citep[e.g.,][]{pbrb17,gk18,mm19,pe20,ac21}. 

As a byproduct of our simulations, we have seen that, at least in the case of FRII jets, where energy is dissipated far from the galactic nucleus, the thermal cooling processes are slow enough to not play a relevant role in global jet evolution at these scales.

We have also shown, in different cases and starting from 2D axisymmetric simulations \citep{pqm11,pmqr14}, that the efficiency of the energetic transfer to the ambient medium depends on the relativistic nature of the jet and on its degree of collimation \citep[see also][for a theoretical explanation]{pmqb17}, and this should be considered in cosmological models including AGN feedback. 

\section*{Acknowledgements}
 Computer simulations have been carried out in the Red Espa\~nola de Supercomputaci\'on (Mare Nostrum and Tirant supercomputers) and in the Servei d'Inform\`atica de la Universitat de Val\`encia.  This work has been supported by the Spanish Ministerio de Ciencia through grants PID2019-105510GB-C31, PID2019-107427GB-C33, and from the Generalitat Valenciana through grant PROMETEU/2019/071. JMM acknowledges further financial support from the Spanish Ministerio de Econom\'{\i}a y Competitividad (grant PGC2018-095984-B-I00). We thank the referee for his/her constructive comments.
 
\section*{Data Availability}
The data underlying this article will be shared on reasonable request to the corresponding author.


\begin{thebibliography}{99}
\bibitem[EHT collaboration(2019)]{eht19a} Event Horizon Telescope Collaboration; Akiyama, K., Alberdi, A., Alef, W., et al., ApJ, 2019, {875}, L1 
\bibitem[Aloy et al.(1999)]{alo99} Aloy, M.A., Ib\'a\~nez, J. M., Mart\'{\i}, J.M., G\'omez, J.-L., M\"uller, E., 1999, ApJ, {523}, 125
\bibitem[Angl\'es-Castillo et al.(2021)]{ac21} Angl\'es-Castillo, A., Perucho, M., Mart\'{\i}, J.M., Laing, R.A., MNRAS, 2021, {500}, 1512
\bibitem[Baczko et al.(2016)]{ba16} Baczko, A.-K., Schulz, R., Kadler, M., et al., 2016, A\&A, {593}, A47
\bibitem[Begelman \& Cioffi(1989)]{bc89} Begelman, M.C., Cioffi, D.F., 1989, ApJ, {345}, L21
\bibitem[Blandford \& Znajek(1977)]{bz77} Blandford, R.D, Znajek, R., 1977, MNRAS, {179}, 433
\bibitem[{Birkinshaw(1984)}]{Birkinshaw1984} Birkinshaw, M. 1984, MNRAS, 208, 887
\bibitem[{Birkinshaw(1991)}]{Birkinshaw1991} Birkinshaw, M. 1991, MNRAS, 252, 505
\bibitem[Barkov \& Khangulyan(2012)]{bk12} Barkov, M., Khangulyan, D., 2012, MNRAS, {421}, 1351
\bibitem[Boccardi et al.(2021)]{bo21} Boccardi, B., Perucho, M., Casadio, C., et al., A\&A, 2021, {647}, A67
\bibitem[Childs et al.(2012)]{visit} Childs H. et al., "High Performance Visualization--Enabling Extreme-Scale Scientific Insight", In Proceedingsof SciDAC 2011, July 2011, http://press.mcs.anl.gov/scidac2011
\bibitem[Croston et al.(2011)]{cro11} Croston, J.H., Hardcastle, M.J., Mingo, B., et al., 2011, ApJ, {734}, L28
\bibitem[Fanaroff \& Riley(1974)]{fr74} Fanaroff B.L., Riley J.M., 1974, MNRAS, {167}, 31
\bibitem[Ghisellini \& Celotti(2001)]{gc01} Ghisellini, G., Celotti, A., 2001, A\&A Letters, {379}, 1
\bibitem[Gitti et al.(2010)]{git10} Gitti, M., O'Sullivan, E., Giacintucci, S., et al., 2010, ApJ, {714}, 75
\bibitem[Gourgouliatos \& Komissarov(2018)]{gk18} Gourgouliatos K.N., Komissarov, S.S., 2018a, Nature Astronomy, {2}, 167
\bibitem[Hada et al.(2011)]{ha11} Hada, K., Doi, A., Kino, M., Nagai, H., Hagiwara, Y., Kawaguchi, N. 2011, Nature, {477}, 185
\bibitem[Hada et al.(2016)]{ha16} Hada, K., Kino, M., Doi, A., et al. 2016, ApJ, {817}, 131
\bibitem[Hardcastle et al.(2002)]{hr02} Hardcastle, M.J., Worrall, D.M., Birkinshaw, M., Laing, R.A., Bridle, A.H. 2002, MNRAS, {334}, 182
\bibitem[Harwood et al.(2016)]{har16} Harwood, J.J., Croston, J.H., Intema, H.T., et al., 2016, MNRAS, {458}, 4443
\bibitem[Harwood et al.(2017)]{har17} Harwood, J.J., Hardcastle, M., Morganti, R., et al., 2017, MNRAS, {469}, 639
\bibitem[Homan el al.(2015)]{ho15} Homan, D.C., Lister, M.L., Kovalev, Y.Y., et al., 2015, ApJ, {798}, 134
\bibitem[Ineson et al.(2017)]{ine17} Ineson, J., Croston, J.H., Hardcastle, M.J., Mingo, B., 2017, MNRAS, {467}, 1586
\bibitem[Jeyakumar \& Saikia(2000)]{js00} Jeyakumar, S., Saikia, D., 2000, MNRAS, 311, 397
\bibitem[Kawakatu, Kino \& Nagai(2009)]{ka09} Kawakatu, N., Kino, M., 2006, MNRAS, {370}, 1513
\bibitem[Kawakatu \& Kino(2006)]{kk06} Kawakatu, N., Kino, M., Nagai, H., 2009, ApJ, {697}, L173
\bibitem[Lister et al.(2009)]{li09} Lister, M.L., Cohen, M.H., Homan, D.C., et al., 2009, AJ, {138}, 1874
\bibitem[Machalsky et al.(2007)]{ma07} Machalski, J., Chyzy, K.T., Stawarz, L., Koziel, D, 2007, A\&A, {462}, 43 
\bibitem[Massaglia et al.(2019)]{ma19} Massaglia, S., Bodo, G., Rossi, P., Capetti, S. Mignone, A., A\&A, {621}, A132
\bibitem[Matsumoto \& Masada(2019)]{mm19} Matsumoto, J., Masada, Y., 2019, MNRAS, {490}, 4271
\bibitem[Matthews et al.(2019)]{ma19} Matthews, J.H., Bell, A.R., Blundell, K.M., Araudo, A.T., 2019, MNRAS, 482, 4303
\bibitem[McKean et al.(2016)]{mk16} McKean, J.P., Godfrey, L.E.H., Vegetti, S., et al., 2016, MNRAS, 463, 3143
\bibitem[McNamara et al.(2005)]{mc05} McNamara, B.R., Nulsen, P.E.J., Wise, M.W., et al., 2005, Nature, {433}, 45
\bibitem[Migliori et al.(2020)]{mig20} Migliori, G., Orienti, M., Coccato, L., Brunetti, G., D’Ammando, F., Mack, K.-H., Prieto, M.A., 2020, MNRAS, 495, 1593
\bibitem[Mignone et al.(2010)]{mi10} Mignone, A., Rossi, P., Bodo, G., Ferrari, A., Massaglia, S., 2010, MNRAS, {402}, 7
\bibitem[Mingo et al.(2019)]{min19} Mingo, B., Croston, J., Hardcastle, M., et al., 2019, MNRAS, {488}, 2701
\bibitem[Mukherjee et al.(2020)]{mu20} Mukherjee, D., Bodo, G., Rossi, P., Mignone, A., Vaidya, B., 2021, MNRAS, {505}, 2267
\bibitem[Mukherjee et al.(2021)]{mu21} Mukherjee, D., Bodo, G., Mignone, A., Rossi, P., Vaidya, B., 2020, MNRAS, {499}, 681
\bibitem[Murgia et al.(1999)]{mu99} Murgia, M., Fanti, C., Fanti, R., Gregorini, L., Klein, U., Mack, K.-H., Vigotti, M., 1999, A\&A, 345, 769
\bibitem[Myasnikov, Zhekov \& Belov(1998)]{mya98} Myasnikov, A. V., Zhekov, S. A., Belov, N. A., 1998, MNRAS, {298}, 1021 
\bibitem[Nishikawa et al.(2021)]{Nishikawa2021} {{Nishikawa}, K., {Du\cb{t}an}, I., {K{\"o}hn}, C. \& {Mizuno}, Y.}, 2021, LRCA, {7}, 1
\bibitem[Nulsen et al.(2005)]{nu05} Nulsen, P.E.J., Hambrick, D.C., McNamara, B.R. et al., 2005, ApJ, {625}, L9
\bibitem[O'Dea et al.(2009)]{od09} O'Dea, C.P., Daly, R.A., Kharb, P., Freeman, K., Baum, S.A., 2016, A\&A, {494}, 471
\bibitem[Perucho(2019)]{pe19} Perucho, M., 2019, Galaxies, {7}, 70
\bibitem[Perucho(2020)]{pe20} Perucho, M., 2020, MNRAS, {494L}, 22 
\bibitem[Perucho et al.(2004)]{ph04} Perucho, M., Hanasz, M., Martí, J.M., Sol, H., 2004, A\&A, {427}, 415
\bibitem[Perucho et al.(2007)]{pe07} Perucho, M., Hanasz, M., Martí, J.M., Miralles, J.A., 2007, PhRvE, {75}, 056312
\bibitem[Perucho \& Mart\'{\i}(2003)]{pm03} Perucho M., Mart\'{\i} J.M., 2003, PASA, {20}, 94
\bibitem[Perucho \& Mart\'{\i}(2007)]{pm07} Perucho M., Mart\'{\i} J.M., 2007, MNRAS, {382}, 526
\bibitem[Perucho et al.(2010)]{pe10} Perucho, M., Mart\'{\i}, J.M., Cela, J.M., Hanasz, M., de la Cruz, R., Rubio, F., 2010, A\&A, {519}, A41 
\bibitem[Perucho, Quilis \& Mart\'{\i}(2011)]{pqm11} Perucho, M., Quilis, V., Mart\'{\i}, J.M., 2011, ApJ, {743}, 42 
\bibitem[Perucho et al.(2014)]{pmqr14} Perucho, M., Mart\'{\i}, J.M., Quilis, V., Ricciardelli, E., 2014, MNRAS, {445}, 1462 
\bibitem[Perucho et al.(2017)]{pbrb17} Perucho, M., Bosch-Ramon, V., Barkov, M., 2017, A\&A, {606}, A40
\bibitem[Perucho et al.(2017)]{pmqb17} Perucho, M., Mart\'{\i}, J.M., Quilis, V., Borja-Lloret, M., 2017, MNRAS, {471L}, 120
\bibitem[Perucho et al.(2019)]{pmq19} Perucho, M., Mart\'{\i}, J.M., Quilis, V., 2019, MNRAS, {482}, 3718 (Paper~I)
\bibitem[Porth(2013)]{po13} Porth, O., 2013, MNRAS, 429, 2482
\bibitem[Rawlings \& Saunders(1991)]{rs91} Rawlings, S., Saunders, R., 1991, Nature, {349}, 138
\bibitem[Rieger \& Levinson(2018)]{rl18} Rieger, F.M., Levinson, A., 2018, Galaxies, 6, 116
\bibitem[Rieger \& Duffy(2019)]{rd19} Rieger, F.M., Duffy, P., 2019, ApJL, 886, 26
\bibitem[Rossi et al.(2020)]{ro20} Rossi, P., Bodo, G., Massaglia, S., Capetti, A., 2020, A\&A, {642}, A69
\bibitem[Scheck et al.(2002)]{sch02} Scheck L., Aloy, M.A., Mart\'{\i}, J.M., G\'omez, J.L., M\"uller, E., 2002, MNRAS, {331}, 615
\bibitem[Scheuer(1974)]{sch74} Scheuer, P.A.G., 1974, MNRAS, {166}, 513
\bibitem[Seo, Kang \& Ryu(2021)]{seo21} Seo, J., Kang, H., Ryu, D., 2021, ApJ, 920, 144
\bibitem[Simionescu et al.(2009)]{si09b} Simionescu, A., Roediger, E., Nulsen, P.E.J., et al., 2009, A\&A, {495}, 721
\bibitem[Snios et al.(2018)]{sni18} Snios, B., Nulsen, P.E.J., Wise, M.W., et al., 2018, ApJ, {855}, 71
\bibitem[Stawarz et al.(2014)]{sta14} Stawarz, \L., Szostek, A., Cheung, C.C., et al., 2014, ApJ, {794}, 164
\bibitem[Synge(1957)]{sy57} Synge, J.L., 1957, The Relativistic Gas, North-Holland publishing Company, Amsterdam
\bibitem[Tchekhovskoy(2015)]{tch15} Tchekhovskoy, A., \emph{Kiloparsec-Scale AGN Jets}, in 'The Formation and Disruption of Black Hole Jets', ASSL, 414, Springer Switzerland, p. 45
\bibitem[Tchekhovskoy, Narayan \& McKinney(2011)]{tch11} Tchekhovskoy, A., Narayan, R., McKinney, J.C., 2011, MNRAS, {418}, 79
\bibitem[Tregillis, Jones \& Ryu(2004)]{tre04} Tregillis, I.L., Jones, T.W., Ryu, D., 2004, ApJ, 601, 778
\bibitem[Vega-Garc\'{\i}a, Perucho \& Lobanov(2019)]{vg19} Vega-Garc\'{\i}a, L., Perucho, M., Lobanov, A.P., 2019, A\&A, 627A, 79
\bibitem[Worrall \& Birkinshaw(2006)]{wb06} Worrall, D.M., Birkinshaw M., 2006, in Alloin D., ed., Lecture Notes in Physics, Berlin Springer Verlag Vol. 693, Physics of Active Galactic Nuclei at all Scales. p. 39
\end{thebibliography}



\appendix
\section{Comparison with the case of dilute IGM within the eBC model}
\label{a:cd}

As said in Sec.~\ref{ss:jd}, we consider the evolution of simulation J1 as divided into two phases, an {\it initial} phase and a {\it multidimensional} one, and use the {\it extended Begelman-Cioffi's} model \citep[eBC;][]{bc89,sch02,pm07,pqm11} to interpret the results. In this model, the advance speed of the bow shock along the axial direction, $v_c$, and the ambient density, $\rho_a$, follow the power laws $v_c \propto t^\alpha$ and $\rho_a \propto r^\beta$, respectively. Parameter $\alpha$ controls the axial expansion rate under both the internal beam processes affecting the jet head propagation, and the density decreasing environment. Parameter $\beta$ regulates the sideways expansion of the global jet structure. The model assumes a complete and instantaneous conversion of the injected jet power into internal energy of the shocked gas and a sideways expansion of the cavity mediated by a strong shock, which is a fair assumption in the case of relativistic outflows \citep[see][]{pmqb17}. According to the eBC model, the width of the shocked region, $R_{\rm BS}$, its aspect ratio, $A_{\rm BS}:= L_{\rm BS}/R_{\rm BS}$, and the cocoon pressure, $P_c$, follow
\begin{equation}
    \quad R_{\rm BS} \propto t^{\frac{2 - \alpha}{4 + \beta}}, \quad A_{\rm BS} \propto t^{\frac{-2 + \alpha}{4 + \beta} + 1 + \alpha}, \quad P_c \propto t^{\frac{2 \alpha - 4}{4 + \beta} - \alpha}.
\end{equation}
For values of $\alpha < 2$ and $\beta > -2$, this model leads to the sideways expansion of the cavity and the decrease of pressure. For a given value of $\beta$, the sideways expansion rate grows and the pressure evolution (typically a drop) becomes slower as $\alpha$ decreases (implying a relative slow down or faster drop of the advance velocity evolution).

%
\begin{table*}  
\begin{center}
{\small
\begin{tabular}{ll | ccccc | ccccc |}\hline
&&&{\bf Phase I}&&&&&{\bf Phase II}&&&\\
&&{$\alpha$}&{$\beta$}&{$P_c$}&{$R_{\rm BS}$}&$A_{\rm BS}$&{$\alpha$}&{$\beta$}&{$P_c$}&{$R_{\rm BS}$}&$A_{\rm BS}$\\
\hline
J1 &  Sim &   $0.13$ & $-1.55$  & $-1.59$  & $0.67$ & $0.46$ & $-0.11$  & $-0.52$  &  $-1.25$ & $0.72$ & $0.17$ \\
   & Model&          &          & $-1.66$  & $0.76$ & $0.37$ &           &          &  $-1.10$ & $0.61$ & $0.28$ \\  \hline
J0 &  Sim &   $0.30$ & $-1.55$  & $-1.74$  & $0.65$ & $0.65$ & $-0.22$  & $-0.52$  &  $-1.20$ & $0.65$ & $0.13$ \\
   & Model&        &        & $-1.69$  & $0.69$ & $0.61$ &       
         &         &   $-1.06$ & $0.64$ & $0.14$\\  \hline
\end{tabular}
}
\caption{Values of the exponents of the power laws determining the time evolution of the cocoon pressure ($P_c$), the transversal size of the shocked region ($R_{\rm BS}$) and the shocked region aspect ratio ($A_{\rm BS}$) during the two phases of the evolution of models J1 (this work) and J0 (Paper~I, for reference). The parameters $\alpha$ and $\beta$ are derived from the simulations. The time dependence of $P_{\rm c}$, $R_{\rm BS}$, and $A_{\rm BS}$ is shown as obtained from the simulation and from the eBC model. The time separating phases I and II in the two simulations is different (2.8 Myrs for J1; 2.0 Myrs for J0).}
\label{tab:tab1}
\end{center}
\end{table*}
%

Table~\ref{tab:tab1} shows the values of the exponents of the power laws for the jet head propagation speed, the transversal size of the shocked region, and its pressure, as a function of time for model J1, obtained both from the simulation and the eBC model. The exponents for model J0 (Paper~I) are also shown for reference. The fact that the values of $\alpha$ for the two phases are so different supports the initial hypothesis of dividing the overall long-term evolution of model J1 in two stages as we did in previous 3D simulations (Paper~I).

Comparing the slopes derived from the simulation and from the model, we conclude that the eBC model properly describes the evolution of the simulation J1 although with poorer accuracy than found for J0. Interestingly, the tendency for a faster sideways expansion and a slower pressure decrease for smaller values of $\alpha$ predicted by the model is fulfilled between simulations J0 and J1 and in both phases. 

Focusing on Phase~I, the main difference in the evolution of jet J1 with respect to J0 is a remarkably smaller $\alpha$ exponent in the former, which implies a milder acceleration phase in J1. The fact that the value of $\alpha$ in model J1, $0.13$ (cf. 0.3 for J0), is close to the one of the axisymmetric simulation J45l of Paper~I, $0.07$, could be an indication that the acceleration mechanism based on the wobbling of the terminal shock is not operating in this case. 

The evolution along Phase~II is characterized by the deceleration of the jet propagation, which reflects in negative values of $\alpha$. The origin of this deceleration should be found in the non-linear growth of the Kelvin-Helmholtz modes and the triggering of the {\it dentist-drill effect}. In the case of model J1, this parameter has a smaller absolute value than in model J0, meaning that the deceleration is milder in the former.

\section{Stability analysis}
\label{a:sa}
The stability analysis of an infinite, relativistic, sheared jet in pressure equilibrium with its environment is described by the following differential equation \citep[see, e.g.,][]{Birkinshaw1984,Birkinshaw1991}: 

\begin{equation}
\begin{matrix}
0 \,=\,\dfrac{d^2 p_1(r)}{dr^2}\, +\, \qquad \qquad \qquad \qquad \qquad \qquad \qquad \qquad \qquad \qquad\\ 
\dfrac{d p_1(r)}{dr} \left\{ \dfrac{1}{r} + \dfrac{2 \gamma_{0}(r) ^2 \dfrac{dv_{0,z}(r)}{dr} \left(k - \dfrac{\omega v_{0,z}(r)}{c^2}\right)}{w-k v_{0,z}(r)} - \dfrac{\dfrac{d\rho_0(r)}{dr}}{\rho_0 + \dfrac{p_0}{c^2}}\right\} + \\
p_1(r) \left\{\gamma_{0}(r)^2 \left[\frac{\rho_0(r) (\omega - k v_{0,z}(r))^2}{\Gamma p_0} - \left(k - \dfrac{\omega v_{0,z}(r)}{c^2}\right)^2\right] - \dfrac{n^2}{r^2}\right\},
\end{matrix}
\label{eq:dif}
\end{equation}
where $r$ and $z$ are the radial and axial coordinates, 
$k$ is the wave-number of the Kelvin-Helmholtz normal modes in the axial direction, $n$ is the azimuthal mode number (the number of oscillations around the circumference of the beam) and $\omega$ the frequency. $\Gamma$ is the adiabatic index, and $p$, $\rho$, $v_{z}$ and $\gamma$ are respectively pressure, rest-mass density, axial flow velocity and flow Lorentz factor. Finally, indices $0$ and $1$ refer to unperturbed and perturbed variables, respectively. The perturbations are assumed to be proportional to $ g(r) \exp{(i(k z - \omega t))}$, with $g(r)$ a function of the jet radius that describes the wave amplitude along coordinate $r$.

The equation is solved by means of the shooting method \citep[see][for details]{pe07}, which finds the roots of the equation for $\omega$ (temporal view) or $k$ (spatial view), by fixing $k$ or $\omega$, respectively. The differential equation is integrated from the boundary condition at the jet axis (given by symmetries), for a given combination of $k$-$\omega$, from the axis to an arbitrary point out of the jet, following the profiles in physical variables, and checking whether this wave pattern satisfies the Sommerfeld condition, i.e., no incoming waves, at an arbitrary position outside of the jet radius. The roots for $\omega$ (temporal view) or $k$ (spatial view) are for each $k$ or $\omega$ are then searched using the M\"uller method. The radial profiles of the unperturbed physical variables are provided to the solver as functions that reproduce the radial profiles observed in the simulations. The solutions of the problem are not very sensitive to small changes in these distributions or in jet parameters, as shown by \citet{vg19}, so we apply approximate profiles of the physical variables. We use the temporal approach to study the growth of unstable modes in time, as explained in the main text. We have chosen to solve the equation for azimuthal wavenumber $n=1$, i.e., for helical modes, and consider the cocoon pressure as the equilibrium pressure of the system. 

In the temporal approach, the real part of the solution is the frequency of the unstable wave, and the imaginary part gives the growth-rate of its amplitude. In the figures, shown in the main text, the former are represented by the upper curves, and the latter are the corresponding curves at the bottom of the image, in $c/R_j$ units. The growth-rates give the $e$-folding times in amplitude and therefore, the larger this value, the faster the growth of the instabilities in the system.

\bsp	
\label{lastpage}
\end{document}